\documentclass[a4paper,fleqn,longmktitle]{cas-sc}

\usepackage[numbers,sort&compress]{natbib}
\usepackage{color}
\usepackage{amsmath}   
\usepackage{float}
\usepackage[title]{appendix}
\usepackage{graphicx}  
\usepackage{caption}
\usepackage{subcaption}
\usepackage{booktabs}
\usepackage{array}
\usepackage{tabulary} 
\usepackage{lineno}
\DeclareCaptionSubType*{figure}

\let\oldequation\equation
\let\oldendequation\endequation
\renewenvironment{equation}{\linenomathNonumbers\oldequation}{\oldendequation\endlinenomath}

\begin{document}

\captionsetup[sub]{
labelformat=simple
}

\title [mode = title]{Spurious currents suppression by accurate difference schemes in multiphase lattice Boltzmann method}

\author{\textcolor[RGB]{0,0,1}{Zhangrong Qin}}
\author{\textcolor[RGB]{0,0,1}{Wenbo Chen}}
\author{\textcolor[RGB]{0,0,1}{Chunyan Qin}}
\author{\textcolor[RGB]{0,0,1}{Xin Xu}}

\author{\textcolor[RGB]{0,0,1}{Binghai Wen}}
\cormark[1]
\ead{oceanwen@gxnu.edu.cn}

\address{Guangxi Key Lab of Multi-Source Information Mining \& Security, Guangxi Normal University, Guilin, 541004, China}
\address{School of Computer Science and Engineering, Guangxi Normal University, Guilin, 541004, China}

\cortext[cor1]{Corresponding author:} 

\begin{abstract}
Spurious currents, which are often observed near a curved interface in the multiphase simulations by diffuse interface methods, are unphysical phenomena and usually damage the computational accuracy and stability. In this paper, the origination and suppression of spurious currents are investigated by using the multiphase lattice Boltzmann method driven by chemical potential. Both the difference error and insufficient isotropy of discrete gradient operator give rise to the directional deviations of nonideal force and then originate the spurious currents. Nevertheless, the high-order finite difference produces far more accurate results than the high-order isotropic difference. We compare several finite difference schemes which have different formal accuracy and resolution. When a large proportional coefficient is used, the transition region is narrow and steep, and the resolution of finite difference indicates the computational accuracy more exactly than the formal accuracy. On the contrary, for a small proportional coefficient, the transition region is wide and gentle, and the formal accuracy of finite difference indicates the computational accuracy better than the resolution. Furthermore, numerical simulations show that the spurious currents calculated in the 3D situation are highly consistent with those in 2D simulations; especially, the two-phase coexistence densities calculated by the high-order accuracy finite difference are in excellent agreement with the theoretical predictions of the Maxwell equal-area construction till the reduced temperature 0.2.
\end{abstract}


\maketitle  

\section{Introduction}

\par{The lattice Boltzmann method (LBM) has been developed as an effective numerical method for simulating multiphase flows in recent years \cite{Shan1993,Qian1997,Chen1998,Xu2006,Aidun2010,Wen2014,Succi2015,Huang2015,Li2016}. The main advantage of LBM lies in its mesoscopic and kinetic nature, which allows it to model some of the microscopic physics that lead to complex behaviors at the macroscopic level, and the liquid-gas interface does not need dynamic reconstruction or tracking. However, the LBM also have some undesirable characteristic in simulating multiphase flow such as the spurious currents, which also occurs in other simulating methods \cite{Lafaurie1994,Tryggvason2001,Luo2015,Harvie2006,Ryu2012,2012Spurious}.}

\par{The spurious currents, also known as parasitic velocities, are a small but finite amplitude circulating flow near the curved interface of the droplet that occurs in some numerical multiphase simulation \cite{Shan2006}. The description of spurious currents is given in \autoref{fig5}(a). From a physical point of view, when the simulation reaches equilibrium, we expect zero fluid velocity everywhere. However, we observed that these spurious currents may persist indefinitely, preventing the system from reaching a true equilibrium, thereby reducing the accuracy of the simulation and may cause instability problems, and it is necessary to understand the origin to reduce or eliminate their effects. }

\par{In the past decades, a number of efforts have been made to explore the origin of spurious currents and reduce or eliminate them \cite{Li2016,Connington2012,Chen2014,Guo2021,Li2021}. Thompson \textit{et al.} \cite{Thompson1999} analyzed the spurious currents through an improved RK LBE \cite{Gunstensen1991} and showed that the magnitude of spurious currents was reduced by a factor of two in comparison with the original model. Lishchuk \textit{et al.} \cite{Lishchuk2003} proposed a lattice Boltzmann algorithm for surface tension to reduce the spurious currents by an order of magnitude. Shan  \cite{Shan2006} argued that insufficient isotropy of the discrete gradient in the interaction force could lead to spurious currents in the pseudopotential LB models and they can be reduced by using high-order isotropic terms in computing the discrete gradient. Later, Sbragaglia \textit{et al.} \cite{Sbragaglia2007} extended the work of Shan \cite{Shan2006} to include more neighbors, thus retaining the higher-order isotropic terms. They showed that the spurious currents can be further reduced when using higher-order isotropic terms to calculate the interaction force, but at the same time, the implementation of boundary conditions will become more complex. Yuan and Schaefer \cite{Yuan2006} found that proper equations of state(EOS) can be increased density ratio and reduced spurious currents. Yu and Fan \cite{Yu2010} analyzed the multiple-relaxation-time (MRT) LB models and found that compared with the single-relaxation-time (SRT) LB models, the MRT LB models can reduce the spurious currents by tuning the adjustable relaxation parameters. Pooley and Furtado \cite{Pooley2008} extended the model of Swift \textit{et al.} \cite{Swift1996} by making a suitable choice of the equilibrium distribution function and isotropic stencil for the derivatives to reduce the spurious currents. Wagner \cite{Wagner2003} pointed out that spurious currents were caused by the inconsistent discretization of the driving forces in LB models and showed that using the potential form of the surface tension force can eliminate the spurious currents to the level of machine accuracy. However, the proposed method must add a small amount of numerical viscosity to maintain the stability of the simulation. Following Wagner’s work, Lee and Fischer \cite{Lee2006} demonstrated that the isotropic discretization and the potential form of surface tension force were necessary to eliminate the spurious currents. }

\par{Nevertheless, a detailed analysis of the force balance equation in LBE was carried out by Guo \textit{et al.} \cite{Guo2011}, they found that no matter whether the pressure-tensor form or the chemical-potential form is used, the force will not be balanced at the discrete lattice level, and resulting in the spurious currents in LBE. To reduce the influence of force imbalance, \textcolor{blue}{Lou and Guo} \cite{Lou2015} proposed a Lax-Wendroff propagation scheme for the LB models. The force imbalance can be adjusted by the Courant-Friedrichs-Levy (CFL) number. However, the computational cost is increased since the time step of this scheme is proportional to the CFL number. \textcolor{blue}{Recently, Guo \cite{Guo2021} conducted a rigorous analysis of the discrete balance equation of LBE, identified the structure of force imbalance caused by the  discretization error, and proposed a well-balanced LBE model by modifying the equilibrium distribution function, which can reduced the magnitude of the spurious currents to the order of machine accuracy.} Li \textit{et al.} \cite{Li2021} also proposed an improved scheme is constructed by modifying the equation of state in standard LBE, through which the discretization of $\nabla \left( {\rho c_{\rm{s}}^2} \right)$ is no longer involved in the force calculation to get a similar achievement.}

\par{The above studies show the causes of spurious currents from different perspectives and provide many methods to reduce or eliminate them in multiphase LB models. However, to the best of our knowledge, no one has ever reported about improving the accuracy of the differential calculation to reduce the error of discrete gradient calculation to suppress the spurious currents. In this paper, we investigate the spurious currents from the perspective of differential calculation error, aiming to introduce a method in the standard streaming-collision procedure to eliminate thermodynamic inconsistencies and suppress spurious currents in the multiphase LB model driven by chemical potential. Specifically, the finite difference scheme with high computational accuracy is used in the nonideal force calculation instead of the traditional isotropic difference to reduce the numerical error in the discrete gradient calculation. In Sec. \ref{Sec.2}, a brief review of the chemical-potential multiphase lattice Boltzmann model with the large density ratio and the finite difference schemes is introduced. In Sec. \ref{Sec.3}, the numerical error of the difference scheme is analyzed, and the causes of the spurious currents are investigated, and then, the numerical simulation validation is carried out. Finally, a brief conclusion is drawn in Sec. \ref{Sec.4}.}

\section{Theory and computational method}\label{Sec.2}
\subsection{Lattice Boltzmann method} \label{Sec.2.1}

\par{LBM is a mesoscopic method for describing fluids, which originated from the lattice gas automata method and kinetic theory \cite{Chen1998,Aidun2010,Succi2015,succi2001lattice,Frisch1986}. The intrinsic mesoscopic properties make it a useful tool for simulating complex flow \cite{Li2016,Chen2014}. The lattice Boltzmann equation (LBE) is a special discrete form of the Boltzmann equation, which is completely discretized in velocity, time, and space \cite{He1997,He1997a}. Several collision operators distinguish variants of the LBE, such as the SRT \cite{Chen1991,Chen1992,Qian1992} model, the entropic model \cite{Karlin1999}, and the two-relaxation-time (TRT) \cite{Ginzburg2008,Ginzburg2008a} and MRT \cite{Lallemand2000,Lallemand2003,Luo2011} models. In the present study, we select the MRT-LBE to obtain better numerical stability and computational efficiency \cite{Luo2011}, which can be written as follows \cite{Lallemand2000,Lallemand2003}:}
\begin{equation}
$$	
{f_i}\left( {{\bf{x}} + {{\bf{e}}_i}\delta t,t + \delta t} \right) - {f_i}({\bf{x}},t) =  - {{\bf{M}}^{ - 1}} \cdot {\bf{S}} \cdot \left[ {{\bf{m}} - {{\bf{m}}^{({\rm{eq}})}}} \right] + {F_i}{\kern 1pt}, 
$$\label{Eq.(1)}
\end{equation}

\begin{figure}
\centering
\includegraphics[scale=0.7]{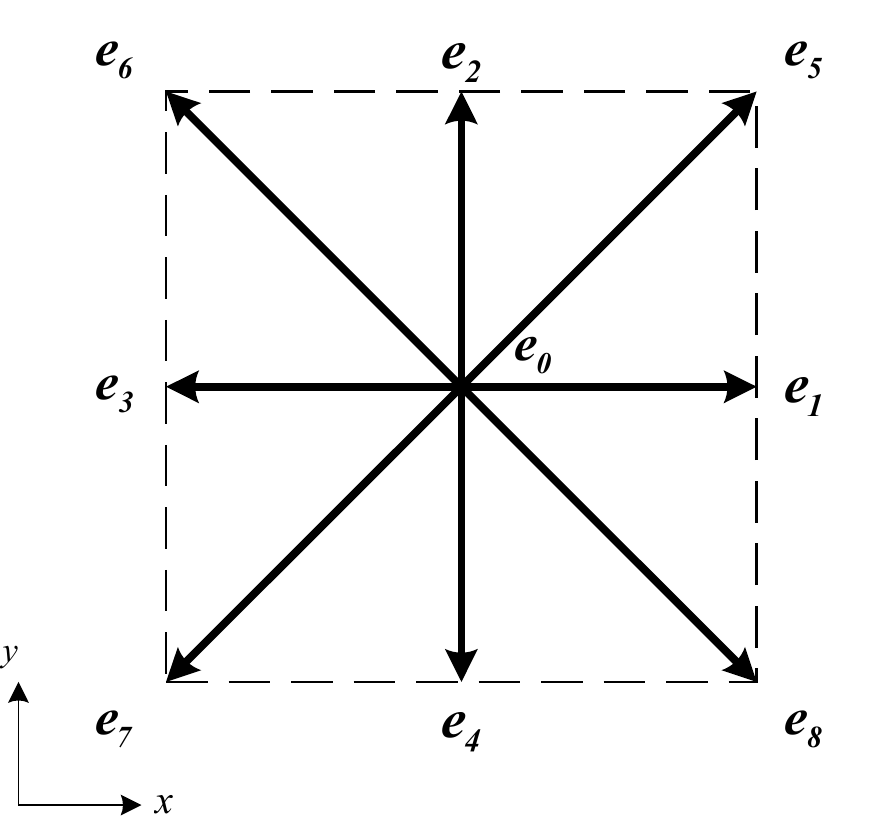}
\caption{Lattice structure and velocity vector of D2Q9 model.}
\label{fig1}	
\end{figure}

\noindent{where ${f_i}({\bf{x}},t)$ is the density distribution function at the lattice site ${\bf{x}}$ and time $t$, ${{\bf{e}}_i}$ is the discrete velocity in the direction of subscript $i$. For the two-dimensional nine-velocity (D2Q9) lattice structure, as shown in \autoref{fig1}, ${{\bf{e}}_i}$ are given by }
\begin{equation}
$$
\begin{array}{l}
{{\bf{e}}_i} = \left\{ {\begin{array}{*{20}{l}}
{(0,0),}&{i = 0,}\\
{c(\cos [\frac{{(i - 1)\pi }}{2}],\quad \sin [\frac{{(i - 1)\pi }}{2}]),}&{i = 1 - 4},\\
{\sqrt 2 c(\cos [\frac{{(2i - 9)\pi }}{4}],\quad \sin [\frac{{(2i - 9)\pi }}{4}]),}&{i = 5 - 8},
\end{array}} \right.\\
\end{array}
$$\label{Eq.(2)}
\end{equation}

\noindent{where $c = {\delta _x}/{\delta _t}$  is the lattice speed, in which ${\delta _x}$ and ${\delta _t}$ are the lattice spacing and time step, respectively; ${\bf{M}}$ is an orthogonal transformation matrix, and ${{\bf{M}}^{{\kern 1pt}  - 1}}$ is the inverse matrix of ${\bf{M}}$. For the D2Q9 model, the transformation matrix ${\bf{M}}$ can be given by \cite{Lallemand2003}}	

\begin{equation}
$$
{\bf{M}} = \left[ {\begin{array}{*{20}{r}}
1&1&1&1&1&1&1&1&1\\
{ - 4}&{ - 1}&{ - 1}&{ - 1}&{ - 1}&2&2&2&2\\
4&{ - 2}&{ - 2}&{ - 2}&{ - 2}&1&1&1&1\\
0&1&0&{ - 1}&0&1&{ - 1}&{ - 1}&1\\
0&{ - 2}&0&2&0&1&{ - 1}&{ - 1}&1\\
0&0&1&0&{ - 1}&1&1&{ - 1}&{ - 1}\\
0&0&{ - 2}&0&2&1&1&{ - 1}&{ - 1}\\
0&1&{ - 1}&1&{ - 1}&0&0&0&0\\
0&0&0&0&0&1&{ - 1}&1&{ - 1}
\end{array}} \right]{\kern 1pt}.
$$\label{Eq.(3)}
\end{equation}

\par{Using the transformation matrix ${\bf{M}}$, the distribution functions ${\bf{f}}$ and its equilibrium distribution function ${{\bf{f}}^{({\rm{eq}})}}$ can be linearly transformed into the moment space:}	
\begin{equation}
	$$
{\bf{m}} = {\bf{M}} \cdot {\bf{f}},\quad {{\bf{m}}^{({\rm{eq}})}} = {\bf{M}} \cdot {{\bf{f}}^{({\rm{eq}})}},\quad {\bf{f}} = {{\bf{M}}^{ - 1}} \cdot {\bf{m}}{\kern 1pt},
	$$\label{Eq.(4)}
\end{equation}

\noindent{where ${\bf{f}} = {\left( {{f_0},{f_1}, \cdots ,{f_8}} \right)^{\rm{T}}}$, ${{\bf{f}}^{{\rm{(eq)}}}}{\rm{ = (}}{f_0}^{(eq)},{f_1}^{(eq)}, \cdot  \cdot  \cdot ,{f_8}^{(eq)}{)^{\rm{T}}}$. ${\bf{m}}$ and ${{\bf{m}}^{{\rm{(eq)}}}}$ represent the velocity moments of the distribution functions ${\bf{f}}$ and their equilibria, respectively. ${\bf{S}}$ is a diagonal matrix composed of relaxation times, and it can be expressed as}
\begin{equation}
$$
{\bf{S}} = {\rm{diag(}}{s_1},{s_2},{s_3},{s_4},{s_5},{s_6},{s_7},{s_8},{s_9}){\kern 1pt},
$$\label{Eq.(5)}
\end{equation}

\noindent{whose elements represent the inverse of the relaxation time for the  distribution function to relax into an equilibrium distribution function in moment space. In this paper, the relaxation times are chosen as follows \cite{McCracken2005}: ${s_1} = {s_4} = {s_6} = 1$, ${s_2} = 1.64$, ${s_3} = 1.54$, ${s_5} = {s_7} = 1.7$, ${s_8} = {s_9} = 1/\tau $, in which $\tau $ is the non-dimensional relaxation time and related to the viscosity by $v = c_s^2(\tau  - 0.5\delta t)$. If we set all relaxation time ${s_i} = 1/\tau $, then the MRT model reduces to the SRT model, \textcolor{blue}{and its the equilibrium distribution function $f_i^{(eq)}$ can be expressed as \cite{Koelman1991}}}
\begin{equation}
$$
f_i^{(eq)}({\bf{x}},t) = {\omega _i}\rho ({\bf{x}},t)\left[ {1 + \frac{{\left( {{{\bf{e}}_i} \cdot {\bf{u}}} \right)}}{{c_s^2}} + \frac{{{{\left( {{{\bf{e}}_i} \cdot {\bf{u}}} \right)}^2}}}{{2c_s^4}} - \frac{{{{({\bf{u}})}^2}}}{{2c_s^2}}} \right]{\kern 1pt},
$$\label{Eq.(6)}
\end{equation}
	
\noindent{where ${\bf{u}}$ is the fluid velocity and ${c_s} = c/\sqrt 3 $ is lattice sound speed. For the D2Q9 model, the weighting coefficients ${\omega _i}$ are given by ${\omega _0}{\rm{ = }}4/9$, ${\omega _{1 - 4}}{\rm{ = 1}}/9$, and ${\omega _{5 - 8}}{\rm{ = 1}}/36$. According to the moment of the density distribution function, the macroscopic density and velocity can be given by \cite{succi2001lattice}}
\begin{equation}
$$
\rho  = \sum\limits_{i = 0}^8 {{f_i}} {\kern 1pt} {\kern 1pt} {\kern 1pt} {\kern 1pt} ,{\kern 5pt} {\kern 5pt} {\kern 5pt} {\kern 5pt} \rho {\bf{u}} = \sum\limits_{i = 0}^8 {{{\bf{e}}_i}} {f_i}{\kern 1pt}.
$$\label{Eq.(7)}
\end{equation}

\par{The external force is brought into LBE through forcing technology \cite{Guo2002}. In this paper, we chose the exact difference method \textcolor{blue}{(EDM)} proposed by Kupershtokh \textit{et al.} to incorporate the nonideal force ${\bf{F}}$ into the LBE \cite{Kupershtokh2004a,Kupershtokh2004b,Kupershtokh2009,Kupershtokh2010}:}
\begin{equation}
$$
{F_i} = f_i^{(eq)}(\rho ,{\bf{u}} + \Delta {\bf{u}}) - f_i^{(eq)}(\rho ,{\bf{u}}){\kern 1pt},
$$\label{Eq.(8)}
\end{equation}

\noindent{where $\Delta {\bf{u}} = \delta t{\bf{F}}/\rho $, the actual fluid velocity is defined as ${\bf{v}} = {\bf{u}} + \delta t{\bf{F}}/(2\rho )$ \cite{Ginzbourg1994}. }

\subsection{Chemical-potential multiphase model}\label{Sec.2.2}
\par{The chemical potential is the partial differential of the Gibbs free energy to the composition \cite{Jamet2002}. For a nonideal fluid system, following the classical capillarity theory of van der Waals, the free energy functional within a gradient-squared approximation is written as \cite{Swift1996,rowlinson2013molecular,Swift1995,Wen2015}}
\begin{equation}
	$$
	\Psi  = \int {\left[ {\psi (\rho ) + \frac{\kappa }{2}|\nabla \rho {|^2}} \right]} d{\bf{x}}{\kern 1pt},
	$$\label{Eq.(9)}
\end{equation}
	
\noindent{where the first term represents the bulk free energy density and the second term describes the contribution from density gradients in an inhomogeneous system, and $\kappa $ is the surface tension coefficient. The general equation of state and chemical potential can be defined by the free energy density \cite{Jamet2002,rowlinson2013molecular,Zheng2006},}
\begin{equation}
$$
{p_0} = \rho \psi '(\rho ) - \psi (\rho ){\kern 1pt},
$$\label{Eq.(10)}
\end{equation}
and
\begin{equation}
	$$
	\mu  = \psi '(\rho ) - \kappa {\nabla ^2}\rho {\kern 1pt}.
	$$\label{Eq.(11)}
\end{equation}

\noindent{Gradients in the chemical potential act as a thermodynamic force on the fluid. With respect to the ideal gas pressure $c_s^2\rho $, the nonideal force can be evaluated by a chemical potential \cite{Wen2017}:}
\begin{equation}
$$
{\bf{F}} =  - \rho \nabla \mu  + c_s^2\nabla \rho {\kern 1pt}.
$$\label{Eq.(12)}
\end{equation}
\noindent{where the nonideal force {\bf{F}}  is incorporated into the LBE through EDM.}

\par{Solving the linear ordinary differential Eq. \eqref{Eq.(10)} gives the general solution of the free energy density,}
\begin{equation}
	$$
	\psi  = \rho (\int {\frac{{{p_0}}}{{{\rho ^2}}}d\rho }  + C) {\kern 1pt},
	$$\label{Eq.(13)}
\end{equation}

\noindent{where $C$ is a constant. When an EOS is selected, substituting Eq. \eqref{Eq.(13)} into Eq. \eqref{Eq.(11)} will solve the relevant chemical potential, and the constant is eliminated. For example, the famous Peng-Robinson (PR) EOS and its chemical potential are,}
\begin{equation}
$$
{p_0} = \frac{{\rho RT}}{{1 - b\rho }} - \frac{{a\alpha (T){\rho ^2}}}{{1 + 2b\rho  - {b^2}{\rho ^2}}} {\kern 1pt},
$$\label{Eq.(14)}
\end{equation}
\noindent{and}
	\begin{equation}
$$
\mu _{}^{{\rm{PR}}} = RT\ln \frac{\rho }{{1 - b\rho }} - \frac{{a\alpha (T)}}{{2\sqrt 2 b}}\ln \frac{{\sqrt 2  - 1 + b\rho }}{{\sqrt 2  + 1 - b\rho }} + \frac{{RT}}{{1 - b\rho }} - \frac{{a\alpha (T)\rho }}{{1 + 2b\rho  - {b^2}{\rho ^2}}} - \kappa {\nabla ^2}\rho {\kern 1pt},
$$\label{Eq.(15)}
\end{equation}

\noindent{where $R$ is the gas constant, \textcolor{blue}{$a$ is the attraction parameter, $b$ is the volume correction parameter,} and the temperature function is $\alpha (T) = {\left[ {1 + \left( {0.37464 + 1.54226\omega  - 0.26992{\omega ^2}} \right)\left( {1 - \sqrt {T/{T_c}} } \right)} \right]^2}$. In our simulations, the parameters are given by $a = 2/49,b = 2/21$, and $R = 1$. The acentric factor $\omega$ is 0.344 for water. The PR EOS is used in the following simulations of this study. To make the numerical results closer to the actual physical properties, we define the reduced variable ${T_r} = T/{T_c}$ and ${\rho _r} = \rho /{\rho _c}$, in which ${T_c}$ is the critical temperature and ${\rho _c}$ is the critical density.}

\par{A proportional coefficient $k$ is introduced to decouple the length unit between the momentum space and the mesh space \cite{Wen2020}, namely $\delta \hat x = k\delta x$. Here the quantities in the mesh space are marked by a superscript. Following dimensional analysis, the chemical potential in the mesh space can be evaluated by}
\begin{equation}
$$
\hat \mu  = {k^2}\psi '(\rho ) - \hat \kappa {\kern 1pt} {\hat \nabla ^2}\rho {\kern 1pt}  {\kern 1pt}.
$$\label{Eq.(16)}
\end{equation}

\par{This approach greatly improves the stability of the chemical-potential multiphase model, and the transformation holds the mathematical equivalence and has no loss of accuracy \cite{Wen2020}.}

\par{\textcolor{blue}{When the multiphase simulation involves a solid surface, the chemical-potential boundary conditon is applied to handle the wettability of the solid surface and the boundary values of the difference scheme. This chemical-potential boundary conditon has been described in the Appendix. }}

\subsection{Finite difference method }\label{Sec.2.3}
\par{In numerical simulations of multiphase flows by LBM, the calculation of the gradient of some characteristic quantities is often involved, and these calculations should be discretized with suitable numerical schemes. In this section, we will briefly introduce three types of numerical schemes used to calculate the gradients, \textcolor{blue}{namely the isotropic finite difference scheme, the explicit finite difference scheme (EFDS), and the compact finite difference scheme (CFDS).}}	
\subsubsection{Explicit finite difference scheme}\label{Sec.2.3.1}
\par{The explicit finite difference is usually to approximate the derivative of the function by linearly combining the values of a function to be differentiated at neighboring points. For simplicity consider, let $f(x)$ be a continuously derivable function on the closed interval $[{x_0},{x_n}]$. The function values at the nodes ${f_i} = f({x_i})$ for $0 \le i \le n$ and the interval is divided into $n$ subintervals on average: ${x_i} = {x_0} + ih$, $h = {\rm{(}}{x_n} - {x_0})/n$. The first-order derivative ${f'_i} = {\left. {\frac{{\partial f}}{{\partial x}}} \right|_{x = {x_i}}}$ at the point $i$ can be written as}
\begin{equation}
$$
{f'_i} = {\left. {\frac{{\partial f}}{{\partial x}}} \right|_{x = {x_i}}} \approx \frac{1}{h}\sum\limits_{m =  - n/2}^{n/2} {{a_m}} {f_{i + m}} + O\left( {{h^p}} \right)  {\kern 1pt},
$$\label{Eq.(17)}
\end{equation}

\noindent{where the first term ${a_m}$ is the coefficients, the last term indicates that the truncation error of the approximation is of order $p$. Several common explicit central difference schemes with different precision are listed:}

\begin{gather}
{f'_i} = \frac{1}{{2h}}\left( { - {f_{i - 1}} + {f_{i + 1}}} \right) + O\left( {{h^2}} \right) {\kern 1pt},\label{Eq.(18)}\\[1mm]
{f'_i} = \frac{1}{{12h}}\left( {{f_{i - 2}} - 8{f_{i - 1}} + 8{f_{i + 1}} - {f_{i + 2}}} \right) + O\left( {{h^4}} \right) {\kern 1pt},\label{Eq.(19)}\\[1mm] 
{f'_i} = \frac{1}{{60h}}\left( { - {f_{i - 3}} + 9{f_{i - 2}} - 45{f_{i - 1}} + 45{f_{i + 1}} - 9{f_{i + 2}} + {f_{i + 3}}} \right) + O\left( {{h^6}} \right)  {\kern 1pt}.\label{Eq.(20)}
\end{gather}

\subsubsection{Compact finite difference scheme}\label{Sec.2.3.2}
\par{Similarly, a widely used formulation for deriving compact schemes is the 7 grid points stencil \cite{Lele1992}. For the function $f(x)$ in Sec. \ref{Sec.2.3.1}, by combining the derivative and function value of the 7 grid points near the discrete point $i$, and the approximate equation of the first-order derivative at the discrete point $i$ can be written as}
\begin{equation}
$$
\beta {f'_{i - 2}} + \alpha {f'_{i - 1}} + {f'_i} + \alpha {f'_{i + 1}} + \beta {f'_{i + 2}} = c\frac{{{f_{i + 3}} - {f_{i - 3}}}}{{6h}} + b\frac{{{f_{i + 2}} - {f_{i - 2}}}}{{4h}} + a\frac{{{f_{i + 1}} - {f_{i - 1}}}}{{2h}}  {\kern 1pt}.
$$\label{Eq.(21)}
\end{equation}

\noindent{Substituting the left hand side and right hand side of Eq. \eqref{Eq.(21)} into the Taylor series expansion, the relationship between parameters $a,{\kern 1pt}b,{\kern 1pt}c,{\kern 1pt}\alpha,{\kern 1pt}\beta$ can be obtained by matching the coefficients of the same order. Furthermore, if $\alpha  = \beta  = 0$, the explicit difference schemes with different precisions can be constructed.}

\par{For simplicity of calculation, we chose $\beta  = 0$, and Eq. \eqref{Eq.(21)} becomes a $\alpha $ family of tridiagonal schemes, which is usually solved by the well-known Thomas algorithm \cite{atkinson2008introduction}. If a further choice of $c = 0$, a $\alpha$ family with fourth-order accuracy schemes can be obtained. Meanwhile, the value of the coefficient  $a,{\kern 1pt}b,{\kern 1pt}\alpha$ is: $a = 2(\alpha  + 2)/3,{\kern 1pt}b = (4\alpha  - 1)/3$. In particular, with $\alpha {\rm{ = 1/4}}$, this $\alpha$ family of tridiagonal schemes becomes the classic Padé scheme:}
\begin{equation}
$$
{f'_i} + \frac{1}{4}({f'_{i + 1}} + {f'_{i - 1}}) = \frac{3}{2}\frac{{{f_{i + 1}} - {f_{i - 1}}}}{{2h}} {\kern 1pt}.
$$\label{Eq.(22)}
\end{equation}
\par{The truncation error of this scheme is listed in Table \ref{table1}. Moreover, with  $\alpha {\rm{ = 1/3}}$, the Eq. \eqref{Eq.(21)} can reach the formally sixth-order accuracy:}
\begin{equation}
$$
{f'_i} + \frac{1}{3}({f'_{i + 1}} + {f'_{i - 1}}) = \frac{{14}}{9}\frac{{{f_{i + 1}} - {f_{i - 1}}}}{{2h}} + \frac{1}{9}\frac{{{f_{i + 2}} - {f_{i - 2}}}}{{4h}} {\kern 1pt}.
$$\label{Eq.(23)}
\end{equation}

\par{If $c \ne 0$ , then a $\alpha $ family of sixth-order tridiagonal schemes can be constructed, and the coefficients $a, {\kern 1pt}b, {\kern 1pt}c, {\kern 1pt}\alpha$ are as follows: $a = (\alpha  + 9)/6$, $b = (32\alpha  - 9)/15$, $c = ( - 3\alpha  + 1)/10$, where $\alpha {\rm{ = 3/8}}$ results in an eighth-order accuracy scheme:}
\begin{equation}
$$
{f'_i} + \frac{3}{8}({f'_{i + 1}} + {f'_{i - 1}}) = \frac{{25}}{{16}}\frac{{{f_{i + 1}} - {f_{i - 1}}}}{{2h}} + \frac{1}{5}\frac{{{f_{i + 2}} - {f_{i - 2}}}}{{4h}} - \frac{1}{{80}}\frac{{{f_{i + 3}} - {f_{i - 3}}}}{{6h}} {\kern 1pt}.
$$\label{Eq.(24)}
\end{equation}

\subsubsection{Isotropic finite difference scheme}\label{Sec.2.3.3}
\par{In the numerical simulation of LBM, the gradient terms are usually evaluated by the isotropic finite difference scheme \cite{Shan2006,Luo2015,Lee2005}. The isotropic finite difference without any directional bias in the lowest order error term and can maintain isotropy in numerical simulations \cite{Kumar2004}. Its first-order derivative and second-order derivative formulas are as follows:}

\begin{subequations}
\begin{align}		
&{\nabla _c}\phi ({\bf{x}}) = \sum\limits_{i = 0}^{n - 1} w \left( {{{\left| {{{\bf{e}}_i}} \right|}^2}} \right){{\bf{e}}_i}\phi \left( {{\bf{x}} + {{\bf{e}}_i}} \right){\kern 1pt},\tag{25a}\label{Eq.(25a)}
\\
&\textcolor{blue}{\nabla _c^2\phi ({\bf{x}}) = 2\sum\limits_{i = 0}^{n - 1} {w\left( {{{\left| {{{\bf{e}}_i}} \right|}^2}} \right)} \left[ {\phi \left( {{\bf{x}} + {{\bf{e}}_i}} \right) - \phi ({\bf{x}})} \right]}{\kern 1pt},\tag{25b}\label{Eq.(25b)}
\end{align}\label{Eq.(25)}
\end{subequations}

\noindent{where $\phi $ is a function of characteristic quantity, such as the density, effective mass, and chemical potential. $w( {{{| {{{\bf{e}}_i}} |}^2}} )$ are the weights. For the commonly used fourth-order isotropic difference, we use the nearest and next-nearest nodes to evaluate gradient terms. For the nearest nodes, which correspond to the nodes on lattice directions 1, 2, 3, and 4 in \autoref{fig1}, the weight $w( {{{| {{{\bf{e}}_i}} |}^2}} )$ in Eq. \eqref{Eq.(25)} is $w\left( 1 \right){\rm{ = }}1/3$, and for the next-nearest one, which corresponds to the nodes on lattice directions 5, 6, 7, and 8, the weight $w( {{{| {{{\bf{e}}_i}} |}^2}} )$ is $w\left( 2 \right){\rm{ = }}1/12$. Some isotropic finite difference introduce more neighborhood nodes to obtain higher isotropy \cite{Shan2006}. Compared with the conventional finite difference schemes, it involves not only the points along the $x$-direction but also points along the $y$-direction \cite{Kumar2004}, which can be regarded as a linear combination of the second-order central difference in different directions.}

\section{Computation and analysis}\label{Sec.3}
\par{In this section, numerical experiments will be conducted. The numerical simulation results are analyzed to find some methods to suppress the spurious currents in the multiphase flow lattice Boltzmann method driven by chemical potential. First, the numerical error of the finite difference method in Sec. \ref{Sec.3.1} will be analyzed and tested via simulations of the flat interface problem. Then the origin of the spurious currents is investigated. Finally, using different finite difference schemes to simulate the two-dimensional circular droplet problem numerically, its spurious currents are analyzed and compared. In the following sections, unless otherwise stated, the two-dimensional circular droplet density field is initialized by \cite{Huang2011}:}
\begin{equation}
	$$\rho (x,y) = \frac{{{\rho _l} + {\rho _g}}}{2} - \frac{{{\rho _l} - {\rho _g}}}{2}\tanh \left[ {\frac{{2\left( {{r} - {r_0}} \right)}}{W}} \right]{\kern 1pt},$$\label{Eq.(26)}
\end{equation}
\noindent{where ${\rho _l}$ and ${\rho _g}$ are the densities of the liquid and vapor phases, $W{\rm{ = }}10$ is the initial interface width, and the circular droplet of radius ${r_0} = 60$ is placed at the center of the computational domain, and $r = \sqrt {{{\left( {x - {x_0}} \right)}^2} + {{\left( {y - {y_0}} \right)}^2}} $, in which $({x_0},{y_0})$ is the center point of the domain. The domain of the simulations is a $300 \times 300$ square with the periodic boundary conditions. }

\begin{table}[]
	\caption{Truncation error in EFDS and CFDS for first derivative calculations.}
	\label{table1} 
	\normalsize
	\centering
	\begin{tabular}{ccc}    
		\toprule   	
		Parameter & \quad\quad  Scheme \quad\quad   & Truncation error \\    
		\midrule   
		$a = 1, {\kern 1pt}\alpha  = \beta  = b = c = 0$ & second-order EFDS & $\frac{1}{{3!}}{h^2}{f^{(3)}}$ \\
		\specialrule{0em}{5pt}{5pt}
		$a = \frac{4}{3}, {\kern 1pt}b =  - \frac{1}{3}, {\kern 1pt}\alpha  = \beta  = c = 0$ &  fourth-order EFDS & $ - \frac{4}{{5!}}{h^4}{f^{(5)}}$ \\ 
		\specialrule{0em}{5pt}{5pt}
		$a = \frac{3}{2}, {\kern 1pt}b =  - \frac{3}{5},{\kern 1pt}c = \frac{1}{{10}}, {\kern 1pt}\alpha  = \beta  = 0$ & sixth-order EFDS & $\frac{{36}}{{7!}}{h^6}{f^{(7)}}$ \\
		\specialrule{0em}{5pt}{5pt}
		$\alpha = \frac{1}{4}, {\kern 1pt}a =\frac{3}{2}, {\kern 1pt} \beta= b = c = 0$ & fourth-order CFDS & $ - \frac{1}{{5!}}{h^4}{f^{(5)}}$ \\
		\specialrule{0em}{5pt}{5pt}
		$\alpha = \frac{1}{3}, {\kern 1pt}a =\frac{14}{9},  {\kern 1pt}b =\frac{1}{9},{\kern 1pt} \beta = c = 0$ & sixth-order CFDS & $\frac{4}{{7!}}{h^6}{f^{(7)}}$ \\
		\specialrule{0em}{5pt}{5pt}
		$\alpha = \frac{3}{8}, {\kern 1pt}a =\frac{25}{16}, {\kern 1pt}b =\frac{1}{5},{\kern 1pt}c =-\frac{1}{80},{\kern 1pt} \beta= 0$ & eighth-order CFDS & $ - \frac{{36}}{{9!}}{h^8}{f^{(9)}}$\\
		\bottomrule   
	\end{tabular}  
\end{table}

\begin{figure}[]
	\centering
	\includegraphics[scale=0.8]{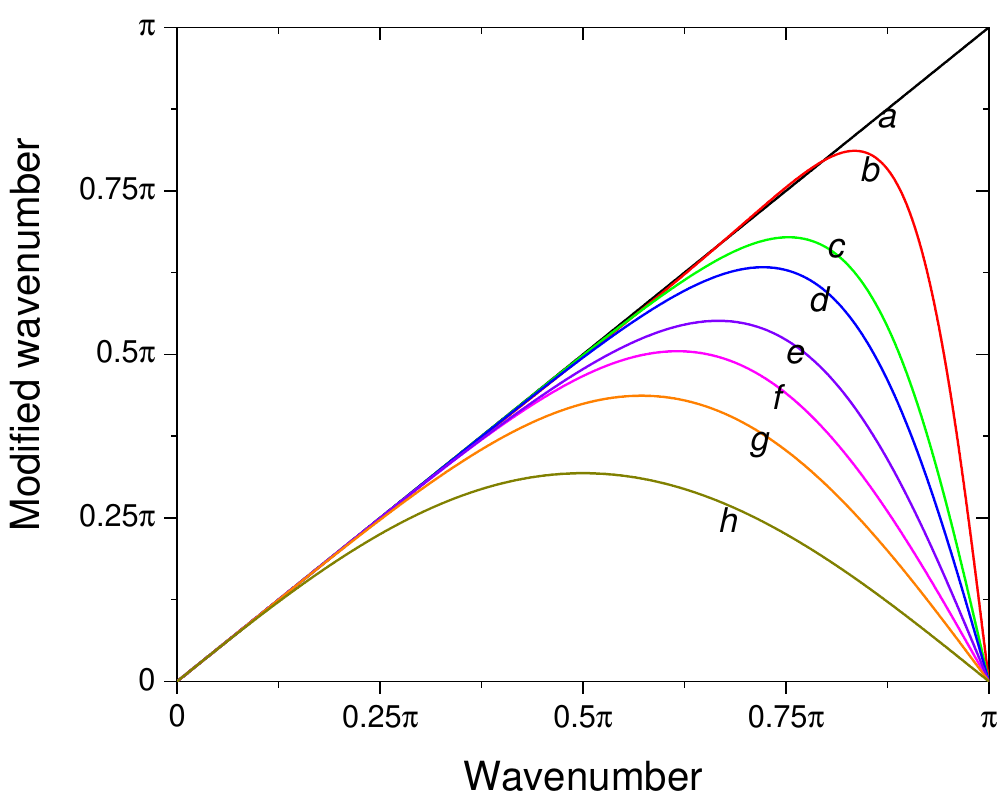}
	\caption{Plot of modified wavenumber vs wavenumber for first derivative approximations: (a) Exact Differentiation, (b) fourth-order OCFDS, (c) eighth-order CFDS, (d) sixth-order CFDS, (e) fourth-order EFDS, (f) sixth-order EFDS, (g) fourth-order EFDS, and (h) second-order EFDS.}
	\label{fig2}
\end{figure}

\subsection{Numerical error analysis of finite difference methods}\label{Sec.3.1}
\par{The analysis of the above chemical-potential models shows that almost all the interparticle interactions include the calculation of gradients and the Laplacian operator, which means that the numerical error in the discrete numerical scheme plays an important part in the numerical simulation. In this section, we will analyze the truncation error and resolution of the difference schemes.}

\par{The truncation error of the EFDS and CFDS are listed in Table \ref{table1}. By comparing the truncation error of the same order difference schemes, it can be concluded that the CFDS can obtain higher accuracy than the EFDS when using the same number points for approximation, such as the truncation error of the fourth-order CFDS is smaller than fourth-order EFDS. Fourier analysis provides an effective way to evaluate the error associated with differencing schemes \cite{vichnevetsky1982fourier}. This method considers the accuracy of the numerical solution of the difference schemes in wave space. It can be used to analyze the resolution of the difference schemes under different scales physical quantities. The Fourier transform of the left and right sides of Eq. \eqref{Eq.(21)} is given as}
\begin{equation}
	$$\begin{array}{l}
\frac{i\omega '}{{h}}[\beta ({e^{ - 2i\omega }} + {e^{2i\omega }}) + \alpha ({e^{ - i\omega }} + {e^{i\omega }}) + 1]\hat f{e^{(iwx/h)}}\\
{\kern 1pt} {\kern 1pt} {\kern 1pt} {\kern 1pt}  = [\frac{c}{{6h}}({e^{3i\omega }} - {e^{ - 3i\omega }}) + \frac{b}{{4h}}({e^{2i\omega }} - {e^{ - 2i\omega }}) + \frac{a}{{2h}}({e^{i\omega }} - {e^{ - i\omega }})]\hat f{e^{(iwx/h)}}{\kern 1pt},
	\end{array}$$\label{Eq.(27)}
\end{equation}

\noindent{where $\omega $ is the wavenumber, $\omega '$ is modified wavenumber. From this equation, the modified wavenumber is derived as:}

\begin{equation}	
	$$\omega '(\omega ) = \frac{{a\sin \omega  + (b/2)\sin 2\omega  + (c/3)\sin 3\omega }}{{1 + 2\alpha \cos \omega  + 2\beta \cos 2\alpha }}.$$\label{Eq.(28)}
\end{equation}
\par{The relation between the modified wavenumber $\omega '$ and wavenumber $\omega $ of the variety for difference schemes is shown in \autoref{fig2}. It can be seen that modified wavenumber of the CFDSs are generally closer to the exact wavenumber (a black line $a$ in \autoref{fig2}) than those of the EFDSs. In other words, the resolution of the CFDSs is better than that of the EFDSs. Ideally, the modified wavenumber is equal to the exact wavenumber. From this, we impose some accuracy constraints to construct a series of high-resolution fourth-order CFDS. Their truncation can be expressed as: $\frac{{12b + 72c - 4\alpha }}{{5!}}{h^4}{f^{(5)}}$, where the difference scheme with the highest resolution is called the optimal fourth-order CFDS (4th-order OCFDS), and its parameters are $a = 1.541$, $b = 0.40667$, $c = -0.0541132$, $\alpha = 0.446776$, and $\beta = 0$. The scheme is drawn in \autoref{fig2} (the red line $b$). Within the same wavenumber range, the resolution of fourth-order OCFDS is better than eighth-order CFDS. That is, the high-resolution difference scheme can also be constructed through a low-order accuracy scheme.}

\begin{figure}[]
\begin{minipage}[t]{0.5\linewidth}
	\centering
	\includegraphics[scale=0.7]{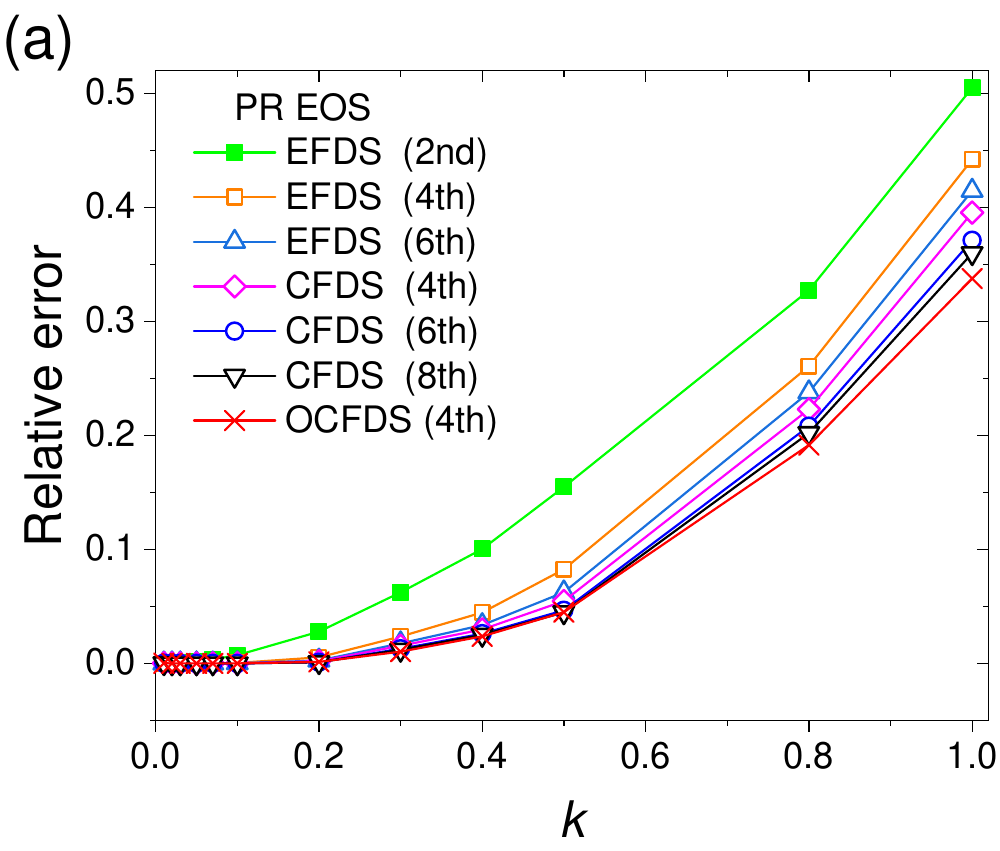}
	\phantomsubcaption	\label{fig3a}
\end{minipage}%
\begin{minipage}[t]{0.5\linewidth}
	\centering
	\includegraphics[scale=0.7]{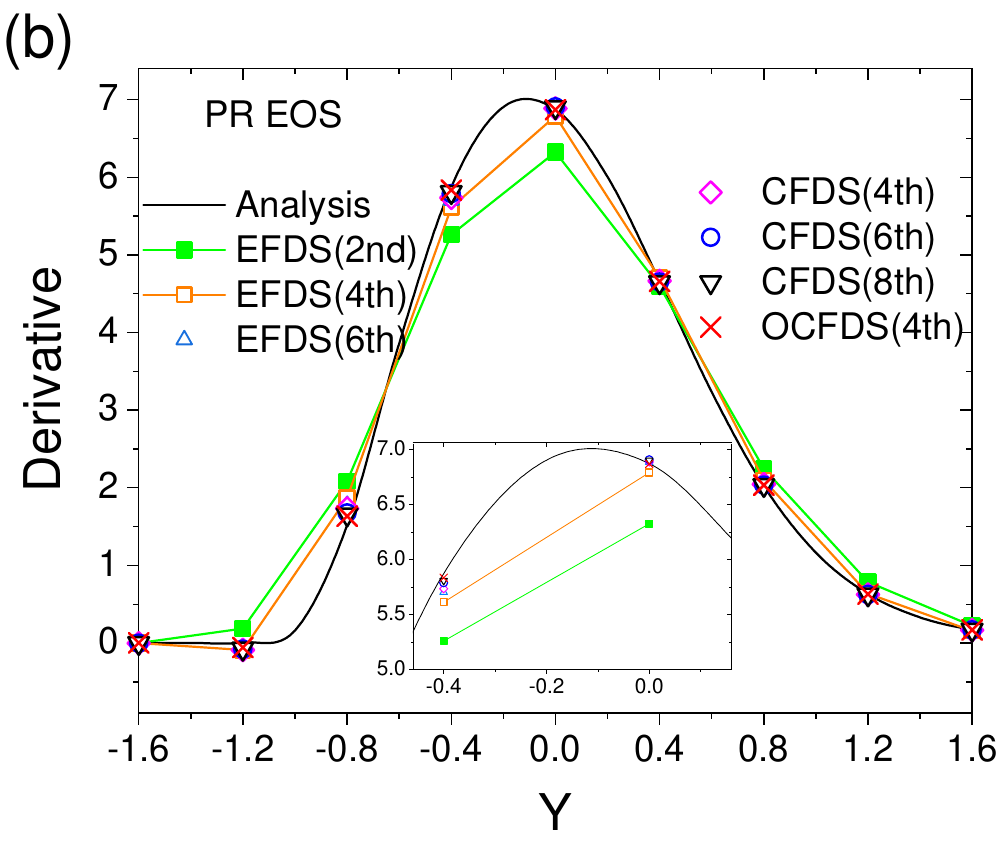}
	\phantomsubcaption\label{fig3b}
\end{minipage}
\caption{(a) Relative errors of the derivative and (b) comparison of the numerical derivatives. The numerical derivatives calculate by the second-, fourth-, sixth-order EFDSs, the fourth-, sixth-, eighth-order CFDSs, and the fourth-order OCFDS.}
\label{fig3}
\end{figure}
\begin{figure}[]
\centering
\includegraphics[scale=0.8]{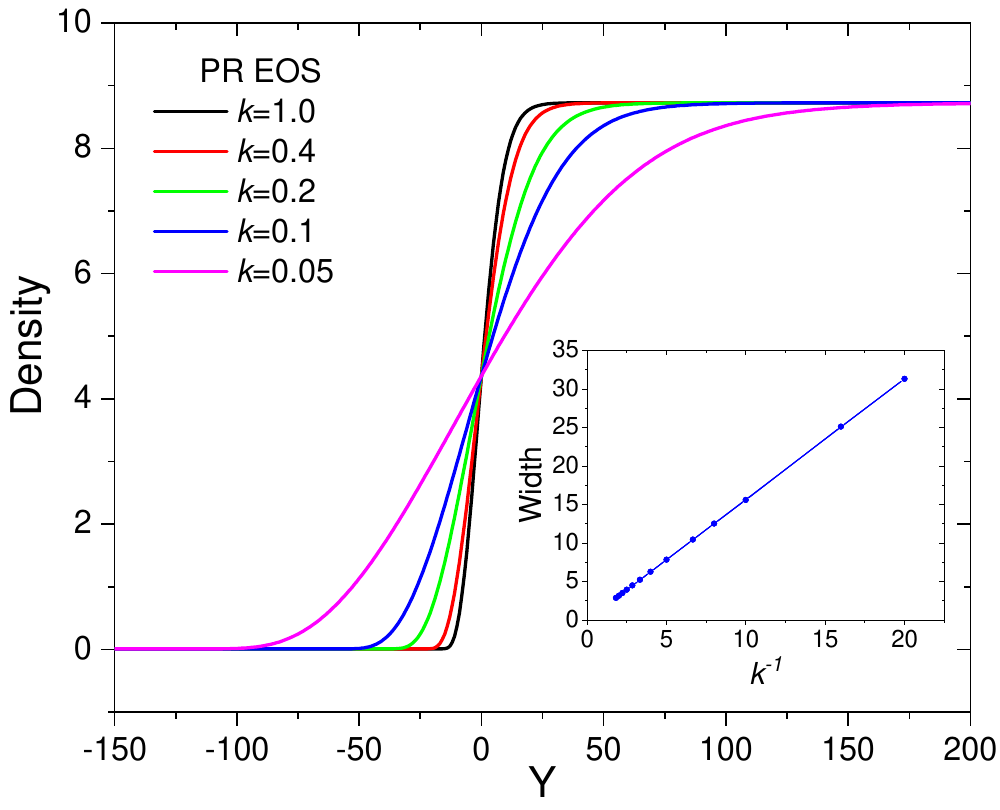}	
\caption{Density profiles for a flat interface of the Chemical-potential model. Inset: the relationship between interface width and proportional coefficient $k$. They have the same interface width in the momentum space, which is equal to the slope.}
\label{fig4}
\end{figure}

\par{In practical calculation, the resolution and truncation order of the finite difference scheme determine the overall error characteristics of the finite difference approximation. To investigate the performances and computational accuracy of different difference methods in this situation, we consider a simple liquid-vapor system with a flat phase interface and expand along the $y$-axis. At the boundary, the liquid and gas phase maintain their bulk density ${\rho _l}$ and ${\rho _g}$, namely, $\rho ( - \infty ) = {\rho _g}$ and $\rho ( + \infty ) = {\rho _l}$. When the system is in equilibrium, the following mechanical equilibrium condition must be satisfied}
\begin{equation}
	$$\nabla  \cdot \mathop {\bf{P}}\limits^ \leftrightarrow  ({\bf{x}}) = 0{\kern 1pt}.$$\label{Eq.(29)}
\end{equation}
\par{For this flat interface problem, the density distribution $\rho (y)$ and its derivatives $\frac{{d\rho }}{{dy}}$ can be easily obtained by solving Eq. \eqref{Eq.(29)}, which will be served as the analytical solutions. The spatial derivative of the analytical density $\rho (y)$ calculated by different finite difference schemes under different proportional coefficient $k$ are shown in \autoref{fig3}(a). It can be seen that the relative errors are related to the accuracy and resolution of the difference scheme, and decrease with proportional coefficient $k$ decreasing. In detail, for larger proportional coefficient $k$, there are fewer grid points in the computational mesh at this time, the resolution of the finite difference scheme indicates the computational accuracy more exactly than the accuracy form, so the relative error of the fourth-order OCFDS is significantly smaller than that of other schemes. Similarly, for smaller proportional coefficient $k$, the grid in the computational mesh is sufficiently dense. In this case, the accuracy of the finite difference scheme mainly determines the computational accuracy of the difference scheme, and thus the error caused by the high-order finite difference scheme will be smaller. \autoref{fig3}(b) shows the comparison of the density derivative solved by the EFDSs and CFDSs with the analytical solution under the proportional coefficient 0.4, and the temperature takes ${T_r} = 0.6$. It can be seen that the derivative calculated by the second-order EFDS deviates significantly from the analytical solution. This can be expected as the derivatives calculated by the second-order EFDS is only related to adjacent nodes, which will inevitably lead to obvious errors in systems with large gas-liquid density ratio and may cause the simulation results to be inconsistent with the actual results. In contrast, high-order CFDS and EFDS use multiple adjacent nodes to calculate the derivative, which reduces the error of derivative calculation, and thus makes the calculation results are basically consistent with the analytical solution, but overall the calculation result of the CFDS is more in agreement with the analytical solution. Therefore, we focus on high-order CFDS in the following simulation. }


\par{Furthermore, we also tested the influence of the proportional coefficient $k$ on the simulation. \autoref{fig4} displays the density distribution at the interface with different proportional coefficient $k$ at the temperature 0.6, where the black line with $k = 1.0$ is consistent with the analytical solution. From the figure it can be observed that all simulations have the same density ratio, but the interface thickness is different. In other words, the proportional coefficient $k$ does not change the liquid-vapor density ratio but only affects the interface thickness, and it can be considered as a means to adjust the interface region without changing the macroscopic physics. The relationship between the proportional coefficient $k$ and the interface width as shown in the inset of \autoref{fig4}. Its slopes are equal to the corresponding widths in the momentum space. It is worth mentioning that the interface in LB simulations usually becomes thinner as the temperature decreases \cite{Li2013}. }

\begin{figure}[]
\begin{minipage}[t]{0.5\linewidth}
	\centering
	\includegraphics[scale=0.7]{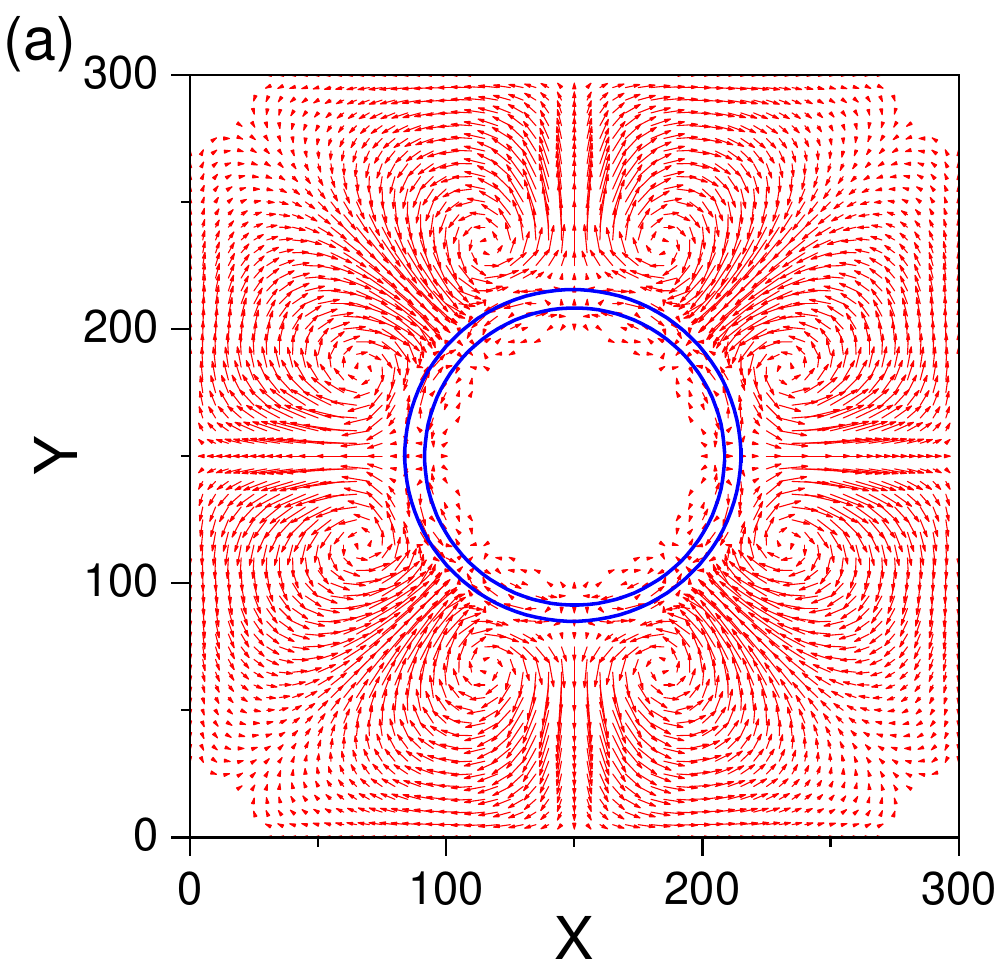}
	\phantomsubcaption	\label{fig5a}
\end{minipage}%
\begin{minipage}[t]{0.5\linewidth}
	\centering
	\includegraphics[scale=0.7]{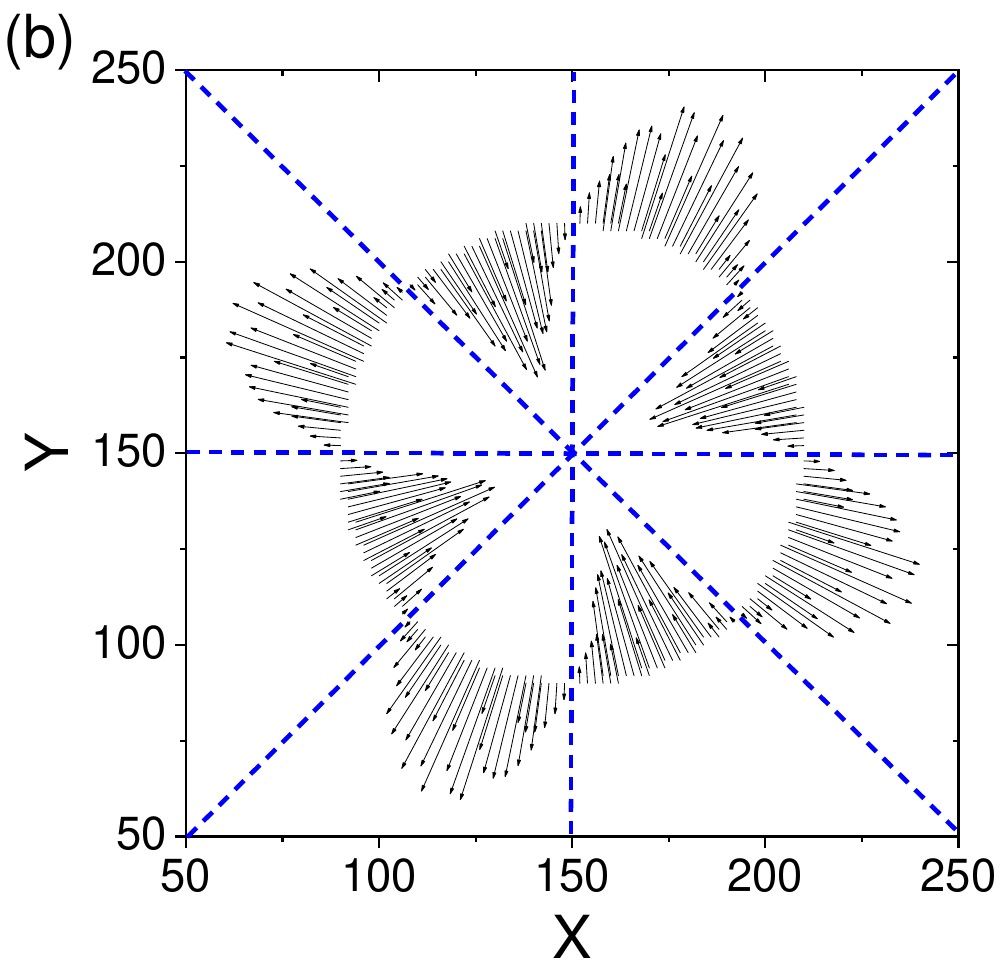}
	\phantomsubcaption\label{fig5b}
\end{minipage}
\caption{\textcolor{blue}{(a) Spurious currents in the 2D simulation of a drop surrounding by gas and (b) the magnitude of the angle deviation of force. }}
\label{fig5}
\end{figure}
\begin{figure}[]
\centering
\includegraphics[scale=0.6]{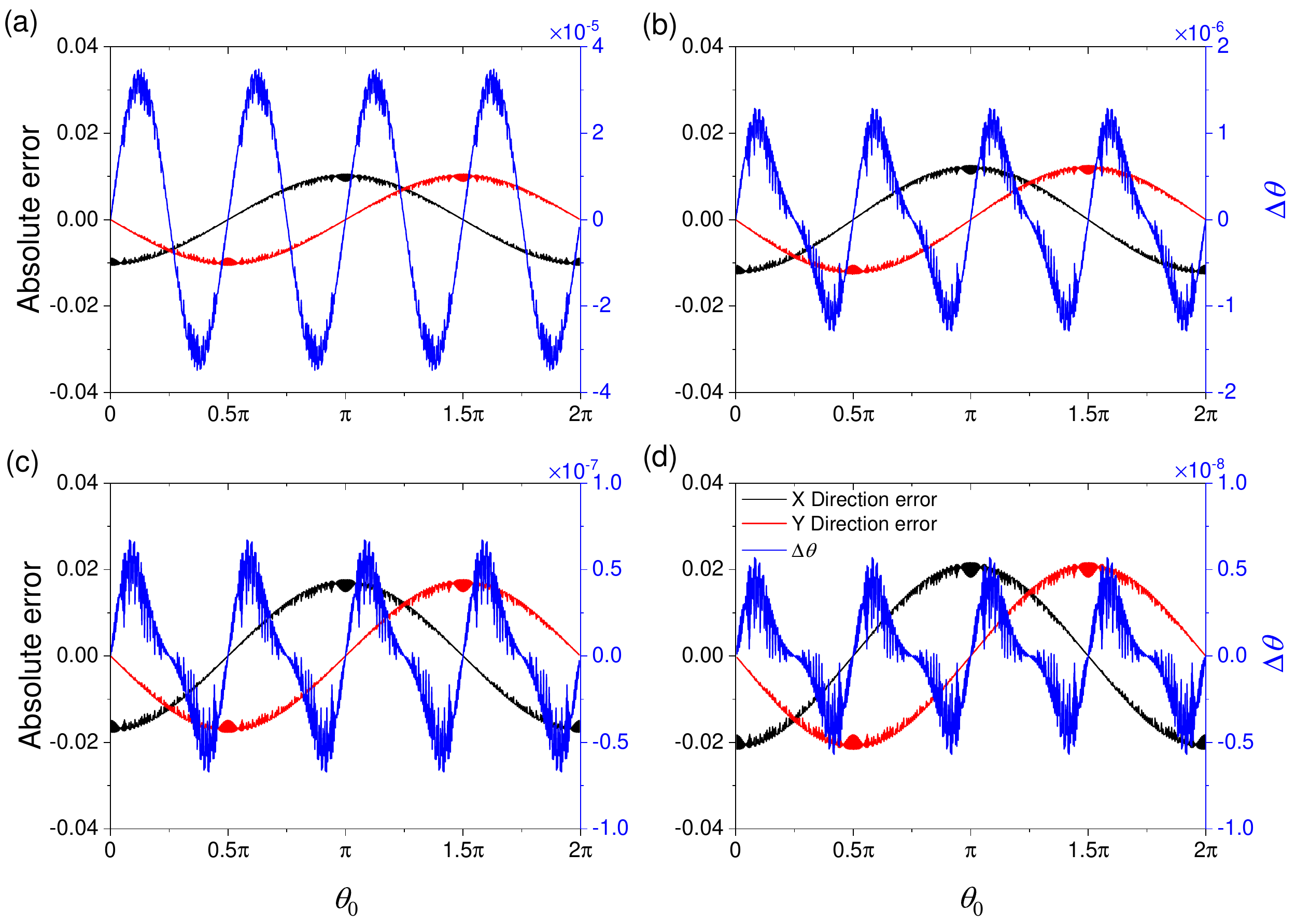}	
\caption{Isotropic finite difference discretization error and angle deviation: (a) fourth-order isotropy, (b) sixth-order isotropy, (c) eighth-order isotropy and (d) tenth-order isotropy.}
\label{fig6}
\end{figure}

\subsection{Nonideal force deviation and spurious currents}\label{Sec.3.2}
\par{In this section, we mainly study the origin of spurious currents. The emergence of the spurious currents is well illustrated by the simulation of a two-dimensional circular droplet without gravity, when the simulation reaches an equilibrium state, eight vortices near the interface of the droplet can be clearly observed as shown in  \autoref{fig5}(a). Ideally, the force produced by the surface tension at the interface of a stable circular droplet should always point towards the center of the droplet \cite{Kruger2017}. Due to the errors in the gradient calculation, the resulting force can be slightly deviated from the center of the droplet and leads to the emergence of spurious currents. To measure the magnitude of the angle deviation of the force from the center of the droplet, we define the following equation:}
\begin{equation}
$$\Delta \theta  = \theta  - {\theta _0} {\rm{ = }} \arctan (\frac{{{F_y}}}{{{F_x}}}) - \arctan \frac{{(y - {y_0})}}{{(x - {x_0})}}{\kern 1pt},$$\label{Eq.(30)}
\end{equation}
\noindent{where the first term is the angle calculated by the nonideal force in the $x$ and $y$ directions, the second term is the original angle, calculated from the position $(x,y)$ and the center point $(x_0,y_0)$.}

\par{In our simulation, the temperature is set to 0.7 and the flow field is initialized using the Eq. \eqref{Eq.(26)}. We calculated the nonideal force near the droplet interface using isotropic scheme and the deviation of the force according to Eq. \eqref{Eq.(30)}, the results are plotted in \autoref{fig5}(b). It shows that the deviation of the force in the horizontal, vertical, and $ \pm 45^\circ $ angles directions have the smallest or even no deviation. The deviation of the force in the approximately $ \pm 22.5^\circ $ and $ \pm 67.5^\circ $ angles directions are largest, which corresponds to the magnitude of spurious currents rotation of in \autoref{fig5}(a). Moreover, we have also studied the force deviations of several high-order isotropic schemes, because the magnitude of spurious currents can be significantly reduced by computing the discrete gradient operator with high-order isotropic discretization schemes \cite{Shan2006,SETA2007}. The results are given in \autoref{fig6}. From the figure, we can see that the numerical errors in the $x$ and $y$ directions gradually increase as the higher-order isotropy is enhanced in the calculation of the discrete gradient operator, but its introduces more neighbor points so that the angle deviation gradually decreases. This is consistent with the above discussion of the spurious currents caused by the angle deviation of force. In the following simulations, we will further verify this view.}

\begin{figure}[]
	\vspace{-1cm}		
	\centering
	\includegraphics[scale=0.6]{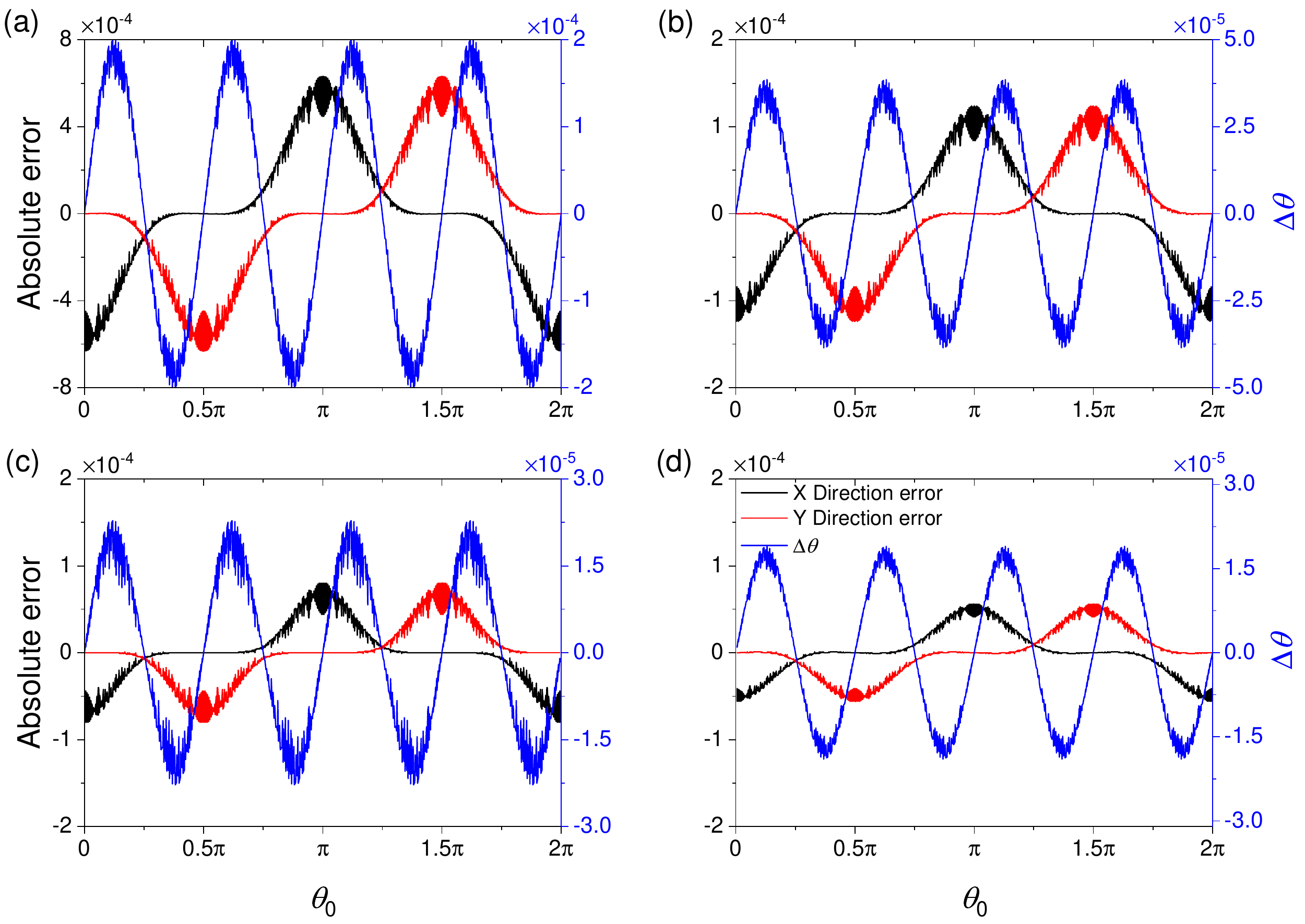}	
	\caption{Discretization error and angle deviation from the CFDS and EFDS: (a) fourth-order EFDS, (b) fourth-order CFDS, (c) sixth-order EFDS, and (d) fourth-order OCFDS.}
	\label{fig7}
\end{figure}

\par{Although the angle deviation of the force can be reduced by high-order isotropic schemes and the spurious currents are reduced to a minimum level, it introduces more neighbor points, which makes the boundary realization more complicated and incurs additional computational costs \cite{Shan2006,Sbragaglia2007}. Moreover, the accuracy of the discrete gradient mainly determines the errors in the numerical calculation of multiphase flow simulations, and the large numerical errors may make the simulation results inaccurate \cite{Qin2018}. For example, it can be clearly seen from \autoref{fig13} that the coexistence densities are given by the isotropic scheme gradually deviates from the analytical curve as the temperature decreases. In order to reduce the numerical errors, we use accurate finite difference instead of isotropic schemes to calculate the nonideal force and the angle deviation of the force as shown in \autoref{fig7}. It is obvious that the angle deviation of the high-order difference scheme is the same level as those of the ${E^{(4)}}$(fourth-order isotropic) scheme, and its numerical error is less than that of the ${E^{(4)}}$. Besides, the angle deviation of the difference scheme decreases as the numerical error decreases. This means that increasing the computational accuracy of the difference scheme can reduce the angle deviation.}

\begin{figure}[pos=H]
\centering
\includegraphics[scale=0.8]{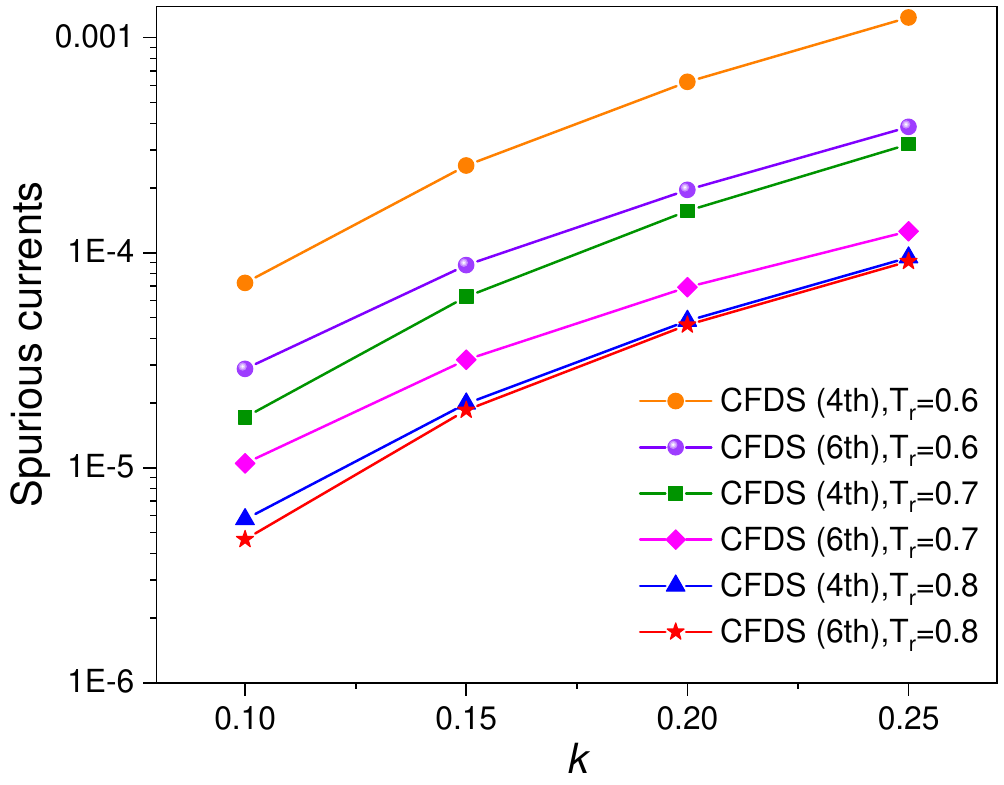}	
\caption{Spurious currents from the fourth-order and sixth-order CFDS are compared at three reduced temperatures. The lower the temperature, the greater the advantage of high-order finite difference scheme.}
\label{fig8}
\end{figure}

\subsection{Spurious current suppression by accurate difference schemes}\label{Sec.3.3}
\par{In this section, we systematically investigate the suppression effect of accurate difference schemes on the spurious currents. In the simulations of two-dimensional circular droplet, the temperatures ${T_r} = 0.6$, 0.7, and 0.8 are used, and the proportional coefficient $k$ changes from 0.1 to 0.25. When the system reaches an equilibrium state, the maximum macroscopic velocity in the domain is regarded as spurious currents. It can be observed from \autoref{fig8} that the lower the temperature, the greater the spurious currents. This is to be expected since the density jump depends on the temperature: the density ratio is larger at lower temperatures, the transition region is steeper, and thus the density jump near the interface is larger. This leads to more numerical errors occur in the discrete gradient calculations and produce higher spurious currents. Therefore, reducing the numerical error in the discrete gradient calculations can suppress the spurious currents, especially for the low temperature (high density ratios) simulation, the suppression effect will be the most obvious.}

\begin{figure}[]
\begin{minipage}[t]{1\linewidth}
	\centering
	\includegraphics[scale=0.8]{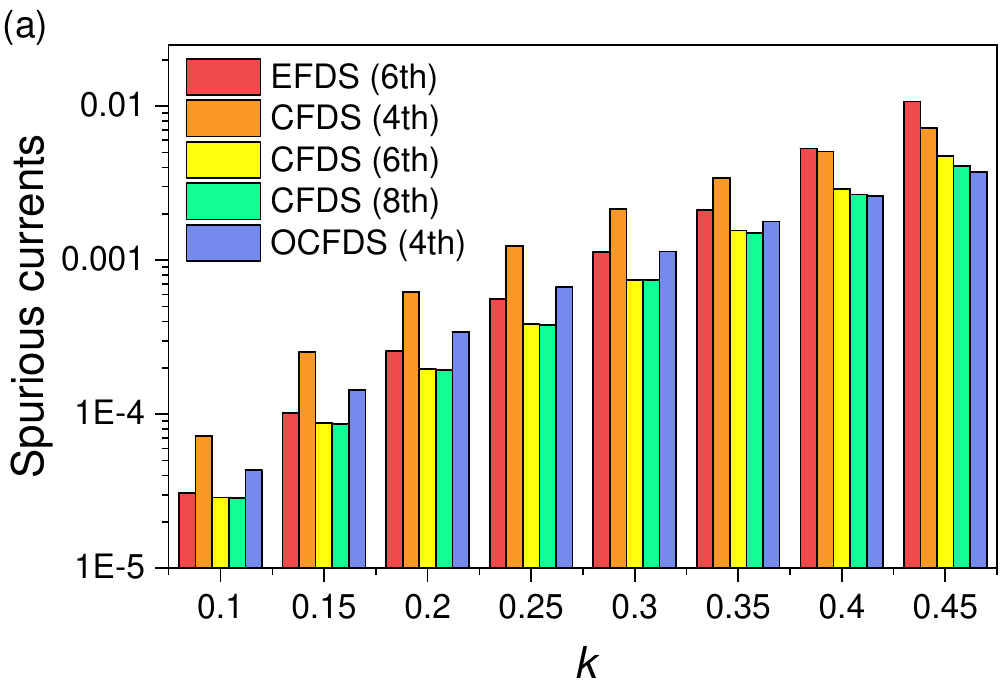}
	\phantomsubcaption	\label{fig9a}
\end{minipage}%
\newline
\begin{minipage}[t]{1\linewidth}
	\vspace{-0.3cm}
	\centering
	\includegraphics[scale=0.8]{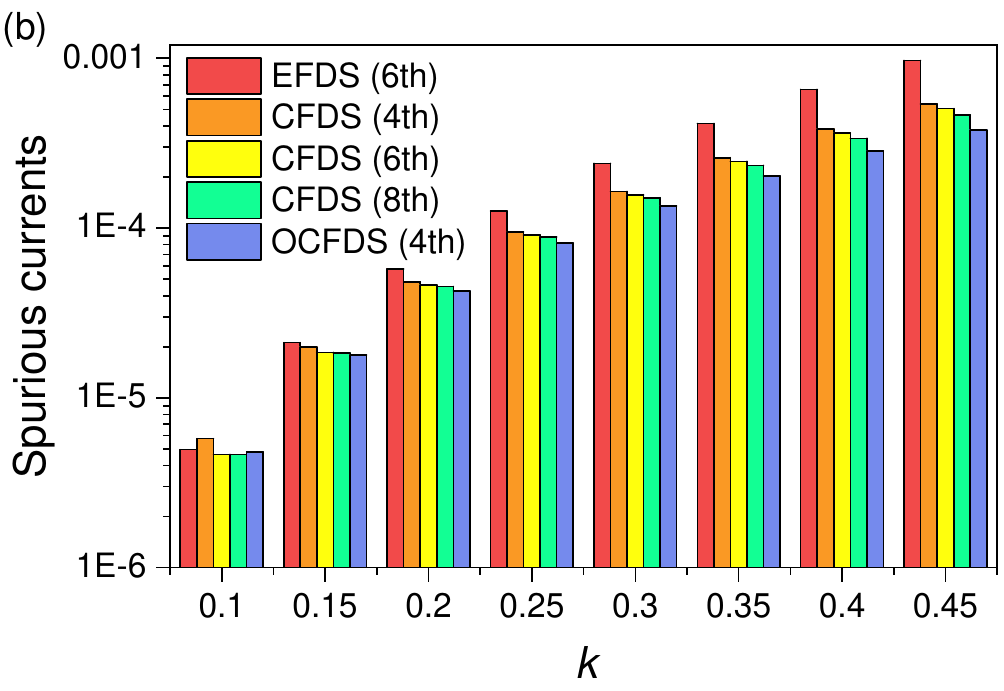}
	\phantomsubcaption\label{fig9b}
\end{minipage}
\caption{\textcolor{blue}{Simulation of circular droplets. Comparison of spurious currents obtained by different finite difference schemes: (a) ${T_r} = 0.6$, (b) ${T_r} = 0.8$. }}
\label{fig9}
\end{figure}

\par{As discussed in the Sec. \ref{Sec.3.1}, when the proportional coefficient $k$ is small, the use of a high-order difference scheme in the computational mesh generates slightly smaller numerical errors. In contrast, when the proportional coefficient $k$ increases, using a high-resolution difference scheme produces smaller numerical errors. To further verify the influence of the numerical error caused by discrete gradient calculations on the spurious currents, we introduced more high-order and high-resolution finite difference schemes in simulation of two-dimensional circular droplets. The spurious currents evaluated by the different finite difference schemes are shown in \autoref{fig9}(a) and (b) for the cases with temperatures of 0.6 and 0.8, respectively. From the figure, we can see that the magnitude of the spurious currents is related to the computational accuracy of the difference scheme and the proportional coefficient $k$. Specifically, for the simulation with a small proportional coefficient, the transition region is smooth and described by dense lattice nodes, the high-order difference schemes have the better effect in suppressing spurious currents compared to the high-resolution difference schemes, particularly in the case of the simulation with temperature 0.6. With the increase of the proportional coefficient, the transition region becomes steeper and is described by fewer lattice nodes, which leads to progressively larger numerical error arising from the calculation of discrete gradients by these difference schemes, and produces larger spurious currents. However, the resolution of the finite difference scheme has a more obvious effect on the computational accuracy at this time, and thus the effect of fourth-order OCFDS in suppressing spurious currents is better than that of other schemes. In brief summary, the accuracy and resolution of finite difference scheme determine its computational accuracy, the numerical error can be reduced by selecting the appropriate finite difference scheme in different simulations, which can effectively suppress the spurious currents.}


\begin{figure}[]
	\centering
	\includegraphics[scale=0.3]{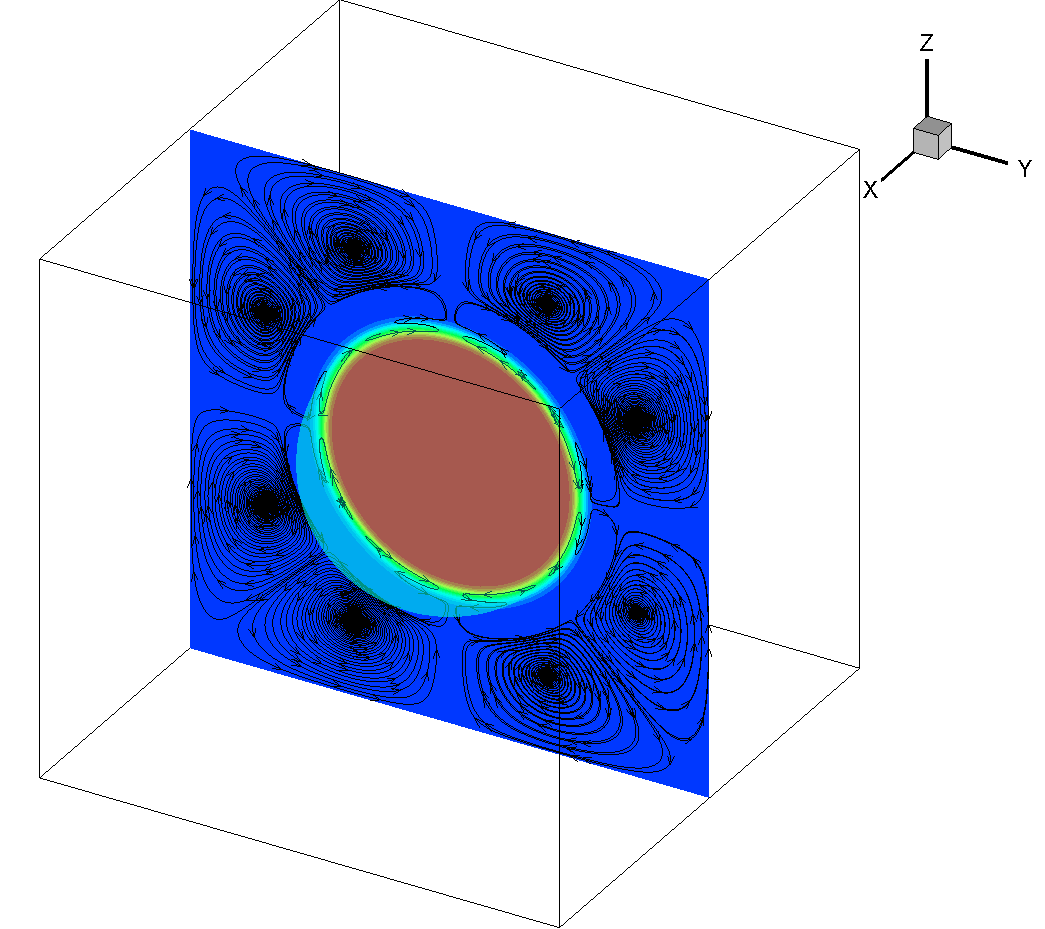}	
	\caption{Spurious currents in the 3D simulation of a drop surrounding by gas. }
	\label{fig10}
\end{figure}

\begin{figure}[pos=H]
	\centering
	\includegraphics[scale=0.8]{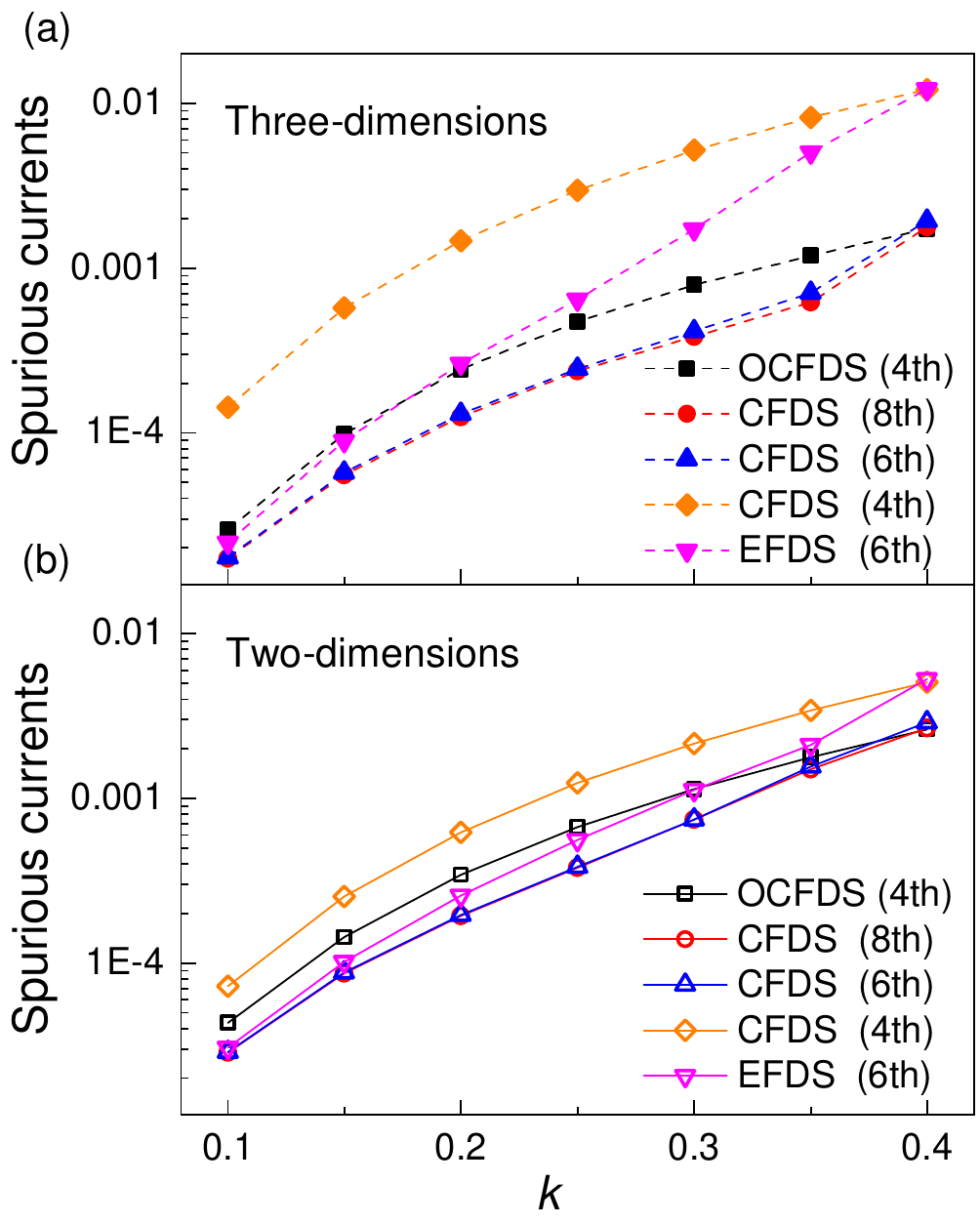}	
	\caption{Spurious currents calculated in (a) 3D simulations are consistent with those in (b) 2D simulations.   }
	\label{fig11}
\end{figure}

\par{\textcolor{blue}{In addition, we investigate the spurious currents in 3D situations. A circular droplet with $r_0=60$ is placed at the center of the $240 \times 240 \times 240$ computational domain, the periodic boundary condition is applied in the $x$, $y$, and $z$ directions, the temperature takes ${T_r} = 0.6$, and the proportional coefficient $k$ changes from 0.1 to 0.4. The spurious currents evaluated by different finite difference schemes are shown in \autoref{fig9}(a), and compared with the two-dimensional results plotted in \autoref{fig9}(b). It can be seen from the figure that when the proportional coefficient $k$ is slightly small, the numerical errors caused by the high-order difference scheme in the gradient calculation is smaller than that of the high-resolution difference scheme, which can suppress the spurious currents better than the high-resolution difference scheme. As the proportional coefficient $k$ increases, these difference schemes generate increasingly larger numerical errors in calculating gradient, and thus produce larger spurious currents. In this case, compared to the high-order difference scheme, the numerical errors caused by the high-resolution difference scheme is relatively small, resulting in smaller spurious currents. For example, the spurious currents from the fourth-order OCFDS is the smallest when $k=0.4$. This is consistent with the two-dimensional conclusion. } }

\begin{figure}[]
	\centering
	\includegraphics[scale=0.8]{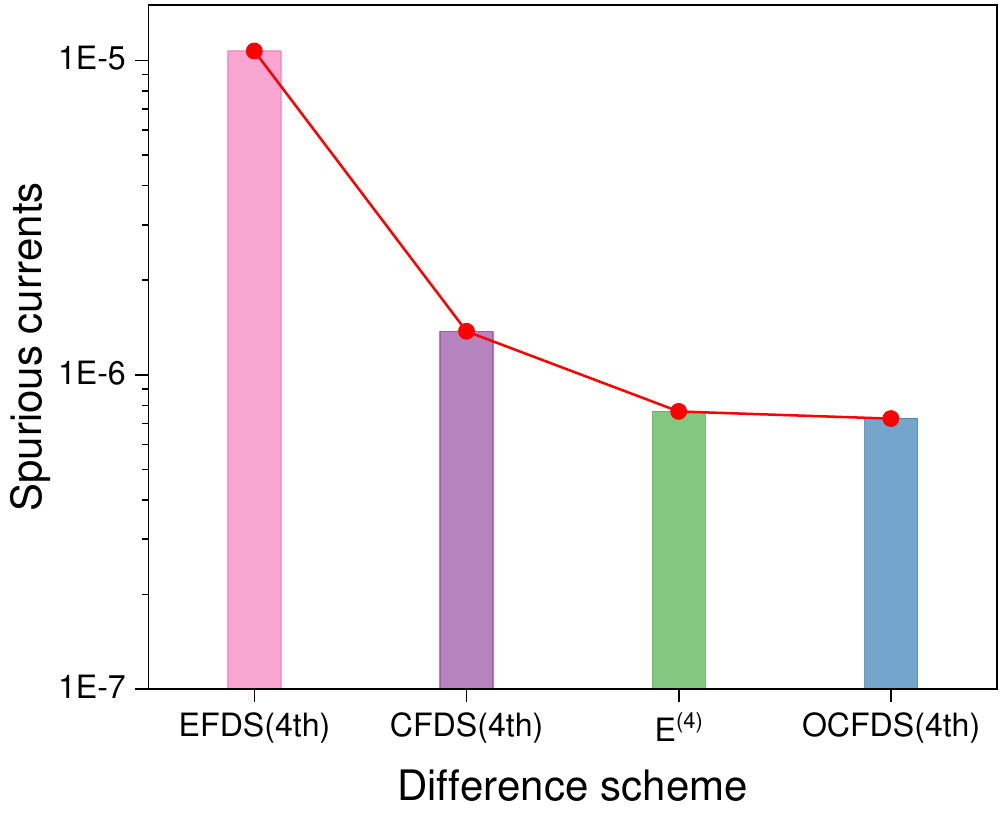}	
	\caption{The spurious currents for the liquid-gas interface system are evaluated by ${E^{(4)}}$, fourth-order EFDS, fourth-order CFDS, and fourth-order OCFDS, respectively. }
	\label{fig12}
\end{figure}

\subsection{Comparison of isotropy and accuracy of difference schemes}\label{Sec.3.4}
\par
\begin{figure}[]
	\centering
	\includegraphics[scale=0.6]{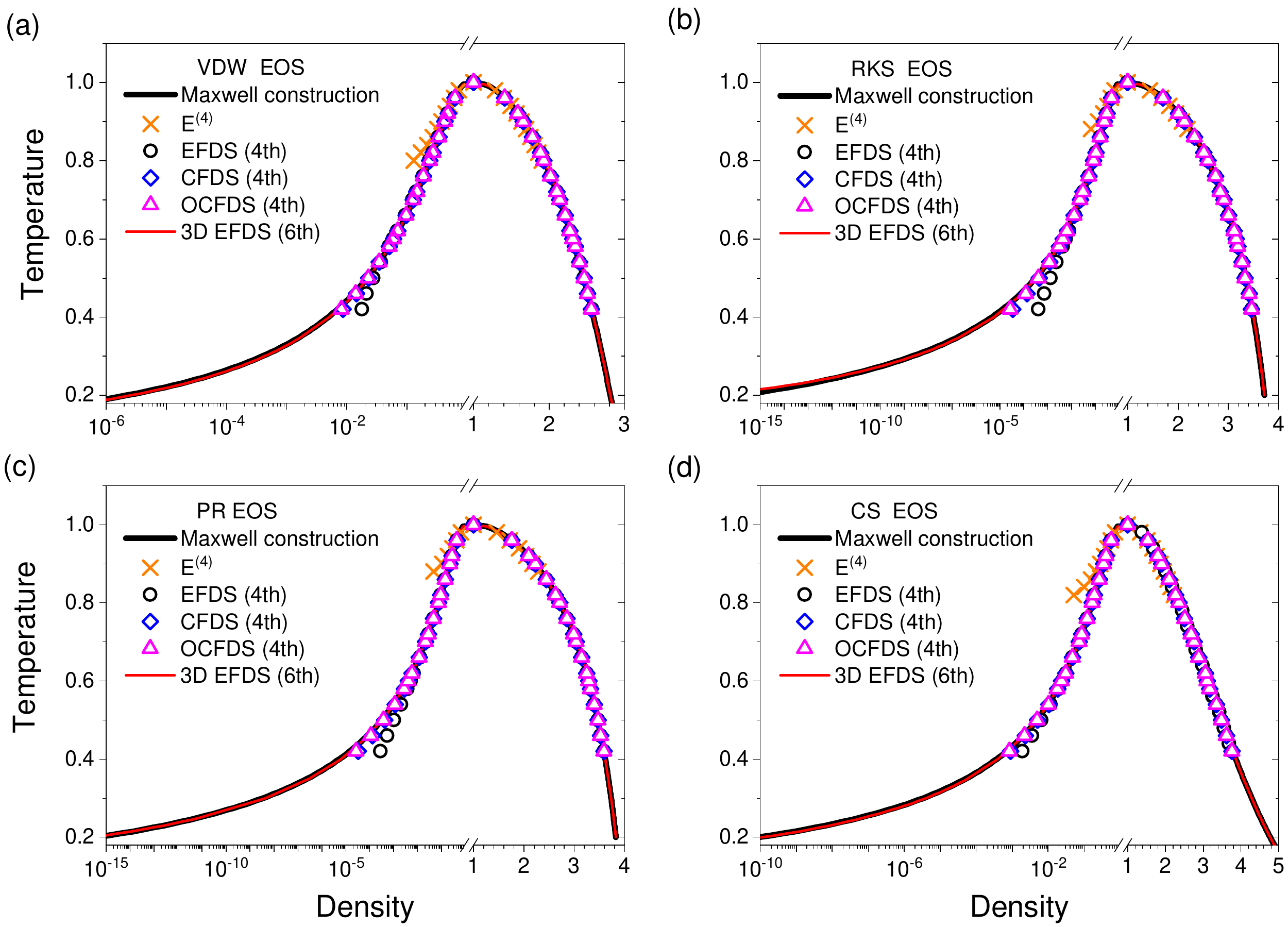}	
	\caption{
		Liquid-gas coexistence densities calculated by the (a) VDW, (b) RKS, (c) PR, and (d) CS EOSs on the chemical-potential multiphase model. The benchmarks are analytically solved by the Maxwell equal-area construction. The gradients are calculated by ${E^{(4)}}$, fourth-order EFDS, fourth-order CFDS, and fourth-order OCFDS in 2D, and sixth-order EFDS in 3D, respectively.
	}
	\label{fig13}
\end{figure}

{In this subsection, a comparison of the isotropic difference scheme and the accurate difference scheme in multiphase flow is performed. The first test problem is a two-dimensional circular droplet. We set the temperature ${T_r} = 0.9$, and the proportional coefficient $k = 0.1$. \autoref{fig12} shows the spurious currents evaluated by the fourth-order EFDS, fourth-order CFDS, fourth-order OCFDS, and ${E^{(4)}}$. Here, it is worthwhile to note that the numerical error of the discrete gradient calculated by ${E^{(4)}}$ is the largest among these difference schemes, but the spurious currents calculated by ${E^{(4)}}$ is an order of magnitude smaller than that of the fourth-order EFDS owing to the nonideal force angle deviation caused by ${E^{(4)}}$ is smaller than that of the fourth-order EFDS. On the other hand, the nonideal force angle deviation caused by fourth-order OCFDS is smaller than that of the ${E^{(4)}}$, and its spurious currents is smaller than that of the ${E^{(4)}}$. This indicates that improving the computational accuracy of the difference scheme to reduce the angular deviation can suppress the spurious currents in the present multiphase flow model. 
\textcolor{blue}{In brief, for the multiphase lattice Boltzmann model, the isotropic finite difference scheme can better suppress spurious currents, but the temperature range of the simulation is limited. For the lower temperatures, the isotropic finite difference may obtain inaccurate simulation results as shown in \autoref{fig13}. While using a high accuracy finite difference scheme to reduce the angular deviation of nonideal force can also effectively suppress the spurious currents, and the temperature range of simulation is larger. } }

\par{The second test problem is a flat interface. We calculated the two-phase coexistence densities in equilibrium state using several difference schemes, and compared with the theoretical values obtained by solving Maxwell equal-area construction equation. In the simulation, the computational domain is set to a height of 400 lattice units, while the width is optional, the central region is initialized as liquid phase, and the rest is filled with the vapor phase. Periodic boundary condition is applied to $x$ and $y$ directions. The relaxation time is 0.8, and the proportional coefficient $k = 0.1$. To systematically investigate the performance of the finite difference schemes in multiphase flow, the tested case covers the usual equations of state, including VDW, RKS, PR, and CS EOSs, whose corresponding bulk free energy density and chemical potential are reported in the literature \cite{Wen2017}. The calculation is performed for 150,000 time steps for each case. The two-phase coexistence densities are shown in \autoref{fig13}.}

\par{Analyzing \autoref{fig13}, it can be clearly seen that the coexistence densities are given by ${E^{(4)}}$ gradually deviating from the analytical curve as the temperature decreases. In contrast, the coexistence densities obtained by the fourth-order EFDS, the fourth-order CFDS, and fourth-order OCFDS agree well with the analytical solution. And with the same accuracy of the difference scheme, the stability range of CFDS at low temperatures with large density ratios is better than that of EFDS. \textcolor{blue}{In addition, we test a flat interface of the two-phase coexistence densities in 3D, where the computational domain is  $400 \times 10 \times 10$, and the periodic boundary condition is applied to all sides. The coexistence densities are given by the sixth-order EFDS, the results are shown in \autoref{fig13} (the red line). It can be observed that the coexistence densities yielded by the sixth-order EFDS is in perfect agreement with the analytical one. 
}
It means that ${E^{(4)}}$ may not be an optimal choice in multiphase flow simulation as its limited accuracies. The high-order finite difference schemes have reliable computational accuracy, which can be an alternative choice, and the higher the accuracy, the larger the stable temperature range of the simulation. Therefore, it can be expected that high-order finite difference scheme may obtain some better results in the low temperature and large density ratio simulations. }

\section{Conclusion}\label{Sec.4}
\par{In this paper, we analyzed the numerical error in the discrete gradient calculation and determined that the deviation of the force caused by the numerical errors will lead to the emergence of spurious currents. To reduce these numerical errors, an accurate finite difference scheme was proposed to evaluate the discrete gradient operator instead of the traditional isotropic difference. Then, the problem of using accurate difference schemes to suppress spurious currents in the multiphase lattice Boltzmann method driven by chemical potential was studied systematically.}
\par{A number of numerical simulations were carried out for validating the proposed scheme. It was shown that using a finite difference scheme with high accuracy can eliminate thermodynamic inconsistencies, and effectively suppress the spurious currents compared to the isotropic scheme. Specifically, the numerical errors in the discrete gradient calculation can be reduced by using a finite difference scheme with high computational accuracy. The two-phase coexistence densities calculated by this scheme are in excellent agreement with the theoretical predictions of the Maxwell equal-area construction till the reduced temperature 0.2. And in the simulation of two-dimensional circular droplets with a density ratio of up to 1000, this scheme can suppress the spurious currents to the order of  ${10^{{\rm{ - }}5}}$. This scheme is also applicable to the 3D situation. In further work, we will focus on achieving a finite difference scheme with higher computational accuracy and use it in more numerical simulations for better simulation results.}

\section*{ACKNOWLEDGMENTS}
This work was supported by the National Natural Science Foundation of China (Grant Nos. 12062005, 11862003, 81860635), the Project of Guangxi Natural Science Foundation (Grant No. 2017GXNSFDA198038, 2021JJA110093), the Graduate Innovation Program of Guangxi Normal University (Grant No. JXXYYJSCXXM-2021-006), Guangxi “Bagui Scholar” Teams for Innovation and Research Project, Guangxi Collaborative Innovation Center of Multisource Information Integration and Intelligent Processing.

\begin{appendices}
	\setcounter{equation}{0}
	\renewcommand{\theequation}{A.\arabic{equation}}
\section{Chemical-potential boundary condition}
 In this appendix, we introduce the wetting boundary condition used in the chemical-potential multiphase lattice Boltzmann method \cite{Wen2017}. Since the present multiphase model is driven by a chemical potential, the implementation of the chemical potential boundary condition is simple and quite natural. The chemical potential can effectively indicate the wettability of a solid surface \cite{Swift1995}. When the EFDSs with the 2nd-, 4th- or 6th- order accuracy are applied to the gradient operator, 1, 2 or 3 layers of solid nodes are required to calculate the gradient of chemical potential on the fluid nodes adjacent to the surface, respectively. To specify the wettability of a solid surface, one can assign a specific chemical potential to these solid nodes. The specific chemical potential gives rise to a chemical-potential gradient between the solid surface and the neighbor fluid, reflecting the nonideal interaction between them and resulting in the wetting phenomena. In order to calculate the density gradient near the boundary, the boundary condition needs to estimate the densities on the solid nodes adjacent to the fluid. Similarly, 1, 2 or 3 layers of solid nodes are required depending on the used accuracy of EFDS. Nevertheless, the densities on these layers of solid nodes are calculated by the weighted average of the nearest neighbor nodes,
\begin{equation}
	$$\rho \left( {{{\bf{x}}_s}} \right) = \frac{{\sum\nolimits_i {{\omega _i}\rho \left( {{{\bf{x}}_s} + {{\bf{e}}_i}{\delta _t}} \right){s_w}} }}{{\sum\nolimits_i {{\omega _i}} {s_w}}}$$
\end{equation}
where ${{x}_{s}}+{{\mathbf{e}}_{i}}{{\delta }_{t}}$ indicates the adjoining nodes, and ${{s}_{w}}$ is a switching function. For the first layer of solid nodes, ${{s}_{w}}=1$ when ${{x}_{s}}+{{\mathbf{e}}_{i}}{{\delta }_{t}}$ is a fluid node; for the second or third layer of nodes, ${{s}_{w}}=1$ when ${{x}_{s}}+{{\mathbf{e}}_{i}}{{\delta }_{t}}$ is in the first or second layer, respectively; otherwise, ${{s}_{w}}=0$.
\end{appendices}

\bibliographystyle{elsarticle-num}
\bibliography{mybib}

\begin{thebibliography}{10}
\expandafter\ifx\csname url\endcsname\relax
  \def\url#1{\texttt{#1}}\fi
\expandafter\ifx\csname urlprefix\endcsname\relax\def\urlprefix{URL }\fi
\expandafter\ifx\csname href\endcsname\relax
  \def\href#1#2{#2} \def\path#1{#1}\fi

\bibitem{Shan1993}
X.~Shan, H.~Chen, {Lattice Boltzmann model for simulating flows with multiple
  phases and components}, Physical Review E 47~(3) (1993) 1815--1819.
\newblock \href {http://dx.doi.org/10.1103/PhysRevE.47.1815}
  {\path{doi:10.1103/PhysRevE.47.1815}}.

\bibitem{Qian1997}
Y.~H. Qian, S.~Chen, {Finite size effect in lattice-BGK models}, International
  Journal of Modern Physics C 8~(4) (1997) 763--771.
\newblock \href {http://dx.doi.org/10.1142/S0129183197000655}
  {\path{doi:10.1142/S0129183197000655}}.

\bibitem{Chen1998}
S.~Chen, G.~D. Doolen, {Lattice Boltzmann method for fluid flows}, Annual
  Review of Fluid Mechanics 30 (1998) 329--364.
\newblock \href {http://dx.doi.org/10.1146/annurev.fluid.30.1.329}
  {\path{doi:10.1146/annurev.fluid.30.1.329}}.

\bibitem{Xu2006}
A.~Xu, G.~Gonnella, A.~Lamura, {Morphologies and flow patterns in quenching of
  lamellar systems with shear}, Physical Review E 74~(1) (2006) 011505.
\newblock \href {http://dx.doi.org/10.1103/PhysRevE.74.011505}
  {\path{doi:10.1103/PhysRevE.74.011505}}.

\bibitem{Aidun2010}
C.~K. Aidun, J.~R. Clausen, {Lattice-Boltzmann method for complex flows},
  Annual Review of Fluid Mechanics 42 (2010) 439--472.
\newblock \href {http://dx.doi.org/10.1146/annurev-fluid-121108-145519}
  {\path{doi:10.1146/annurev-fluid-121108-145519}}.

\bibitem{Wen2014}
B.~Wen, C.~Zhang, Y.~Tu, C.~Wang, H.~Fang, {Galilean invariant fluid-solid
  interfacial dynamics in lattice Boltzmann simulations}, Journal of
  Computational Physics 266~(6) (2014) 161--170.
\newblock \href {http://dx.doi.org/10.1016/j.jcp.2014.02.018}
  {\path{doi:10.1016/j.jcp.2014.02.018}}.

\bibitem{Succi2015}
S.~Succi, {Lattice Boltzmann 2038}, Europhysics Letters 109~(5) (2015) 50001.
\newblock \href {http://dx.doi.org/10.1209/0295-5075/109/50001}
  {\path{doi:10.1209/0295-5075/109/50001}}.

\bibitem{Huang2015}
H.~Huang, M.~C. Sukop, X.~Y. Lu, {Multiphase Lattice Boltzmann Methods: Theory
  and Application}, Wiley, New York, 2015.
\newblock \href {http://dx.doi.org/10.1002/9781118971451}
  {\path{doi:10.1002/9781118971451}}.

\bibitem{Li2016}
Q.~Li, K.~H. Luo, Q.~J. Kang, Y.~L. He, Q.~Chen, Q.~Liu, {Lattice Boltzmann
  methods for multiphase flow and phase-change heat transfer}, Progress in
  Energy and Combustion Science 52 (2016) 62--105.
\newblock \href {http://dx.doi.org/10.1016/j.pecs.2015.10.001}
  {\path{doi:10.1016/j.pecs.2015.10.001}}.

\bibitem{Lafaurie1994}
B.~Lafaurie, C.~Nardone, R.~Scardovelli, S.~Zaleski, G.~Zanetti, {Modelling
  merging and fragmentation in multiphase flows with SURFER}, Journal of
  Computational Physics 113~(1) (1994) 134--147.
\newblock \href {http://dx.doi.org/10.1006/jcph.1994.1123}
  {\path{doi:10.1006/jcph.1994.1123}}.

\bibitem{Tryggvason2001}
G.~Tryggvason, B.~Bunner, A.~Esmaeeli, D.~Juric, N.~Al-Rawahi, W.~Tauber,
  J.~Han, S.~Nas, Y.~J. Jan, {A front-tracking method for the computations of
  multiphase flow}, Journal of Computational Physics 169~(2) (2001) 708--759.
\newblock \href {http://dx.doi.org/10.1006/jcph.2001.6726}
  {\path{doi:10.1006/jcph.2001.6726}}.

\bibitem{Luo2015}
J.~Luo, X.~Y. Hu, N.~A. Adams, {A conservative sharp interface method for
  incompressible multiphase flows}, Journal of Computational Physics 284~(C)
  (2015) 547--565.
\newblock \href {http://dx.doi.org/10.1016/j.jcp.2014.12.044}
  {\path{doi:10.1016/j.jcp.2014.12.044}}.

\bibitem{Harvie2006}
D.~J. Harvie, M.~R. Davidson, M.~Rudman, {An analysis of parasitic current
  generation in Volume of Fluid simulations}, Applied Mathematical Modelling
  30~(10) (2006) 1056--1066.
\newblock \href {http://dx.doi.org/10.1016/j.apm.2005.08.015}
  {\path{doi:10.1016/j.apm.2005.08.015}}.

\bibitem{Ryu2012}
S.~Ryu, S.~Ko, {A comparative study of lattice Boltzmann and volume of fluid
  method for two-dimensional multiphase flows}, Nuclear Engineering and
  Technology 44~(6) (2012) 623--638.
\newblock \href {http://dx.doi.org/10.5516/NET.02.2011.025}
  {\path{doi:10.5516/NET.02.2011.025}}.

\bibitem{2012Spurious}
S.~Zahedi, M.~Kronbichler, G.~Kreiss, {Spurious currents in finite element
  based level set methods for two-phase flow}, International Journal for
  Numerical Methods in Fluids 69~(9) (2012) 1433--1456.
\newblock \href {http://dx.doi.org/10.1002/fld.2643}
  {\path{doi:10.1002/fld.2643}}.

\bibitem{Shan2006}
X.~Shan, {Analysis and reduction of the spurious current in a class of
  multiphase lattice Boltzmann models}, Physical Review E 73~(4) (2006) 47701.
\newblock \href {http://dx.doi.org/10.1103/PhysRevE.73.047701}
  {\path{doi:10.1103/PhysRevE.73.047701}}.

\bibitem{Connington2012}
K.~Connington, T.~Lee, {A review of spurious currents in the lattice Boltzmann
  method for multiphase flows}, Journal of Mechanical Science and Technology
  26~(12) (2012) 3857--3863.
\newblock \href {http://dx.doi.org/10.1007/s12206-012-1011-5}
  {\path{doi:10.1007/s12206-012-1011-5}}.

\bibitem{Chen2014}
L.~Chen, Q.~Kang, Y.~Mu, Y.~L. He, W.~Q. Tao, {A critical review of the
  pseudopotential multiphase lattice Boltzmann model: Methods and
  applications}, International Journal of Heat and Mass Transfer 76~(0) (2014)
  210--236.
\newblock \href {http://dx.doi.org/10.1016/j.ijheatmasstransfer.2014.04.032}
  {\path{doi:10.1016/j.ijheatmasstransfer.2014.04.032}}.

\bibitem{Guo2021}
Z.~Guo, {Well-balanced lattice Boltzmann model for two-phase systems}, Physics
  of Fluids 33~(3) (2021) 31709.
\newblock \href {http://dx.doi.org/10.1063/5.0041446}
  {\path{doi:10.1063/5.0041446}}.

\bibitem{Li2021}
Q.~Li, Y.~Yu, R.~Z. Huang, {Achieving thermodynamic consistency in a class of
  free-energy multiphase lattice Boltzmann models}, Physical Review E 103~(1)
  (2021) 13304.
\newblock \href {http://dx.doi.org/10.1103/PhysRevE.103.013304}
  {\path{doi:10.1103/PhysRevE.103.013304}}.

\bibitem{Thompson1999}
S.~P. Thompson, I.~Halliday, C.~M. Care, {Mesoscopic hydrodynamics of diphasic
  lattice Bhatnagar Gross Krook fluid interfaces}, Physical Chemistry Chemical
  Physics 1~(9) (1999) 2183--2190.
\newblock \href {http://dx.doi.org/10.1039/a809793c}
  {\path{doi:10.1039/a809793c}}.

\bibitem{Gunstensen1991}
A.~K. Gunstensen, D.~H. Rothman, S.~Zaleski, G.~Zanetti, {Lattice Boltzmann
  model of immiscible fluids}, Physical Review A 43~(8) (1991) 4320--4327.
\newblock \href {http://dx.doi.org/10.1103/PhysRevA.43.4320}
  {\path{doi:10.1103/PhysRevA.43.4320}}.

\bibitem{Lishchuk2003}
S.~V. Lishchuk, C.~M. Care, I.~Halliday, {Lattice Boltzmann algorithm for
  surface tension with greatly reduced microcurrents}, Physical Review E 67~(3)
  (2003) 5.
\newblock \href {http://dx.doi.org/10.1103/PhysRevE.67.036701}
  {\path{doi:10.1103/PhysRevE.67.036701}}.

\bibitem{Sbragaglia2007}
M.~Sbragaglia, R.~Benzi, L.~Biferale, S.~Succi, K.~Sugiyama, F.~Toschi,
  {Generalized lattice Boltzmann method with multirange pseudopotential},
  Physical Review E 75~(2) (2007) 26702.
\newblock \href {http://dx.doi.org/10.1103/PhysRevE.75.026702}
  {\path{doi:10.1103/PhysRevE.75.026702}}.

\bibitem{Yuan2006}
P.~Yuan, L.~Schaefer, {Equations of state in a lattice Boltzmann model},
  Physics of Fluids 18~(4) (2006) 42101.
\newblock \href {http://dx.doi.org/10.1063/1.2187070}
  {\path{doi:10.1063/1.2187070}}.

\bibitem{Yu2010}
Z.~Yu, L.~S. Fan, {Multirelaxation-time interaction-potential-based lattice
  Boltzmann model for two-phase flow}, Physical Review E 82~(4) (2010) 046708.
\newblock \href {http://dx.doi.org/10.1103/PhysRevE.82.046708}
  {\path{doi:10.1103/PhysRevE.82.046708}}.

\bibitem{Pooley2008}
C.~M. Pooley, K.~Furtado, {Eliminating spurious velocities in the free-energy
  lattice Boltzmann method}, Physical Review E 77~(4) (2008) 46702.
\newblock \href {http://dx.doi.org/10.1103/PhysRevE.77.046702}
  {\path{doi:10.1103/PhysRevE.77.046702}}.

\bibitem{Swift1996}
M.~R. Swift, E.~Orlandini, W.~R. Osborn, J.~M. Yeomans, {Lattice Boltzmann
  simulations of liquid-gas and binary fluid systems}, Physical Review E 54~(5)
  (1996) 5041--5052.
\newblock \href {http://dx.doi.org/10.1103/PhysRevE.54.5041}
  {\path{doi:10.1103/PhysRevE.54.5041}}.

\bibitem{Wagner2003}
A.~J. Wagner, {The origin of spurious velocities in lattice Boltzmann},
  International Journal of Modern Physics B 17~(1-2) (2003) 193--196.
\newblock \href {http://dx.doi.org/10.1142/s0217979203017448}
  {\path{doi:10.1142/s0217979203017448}}.

\bibitem{Lee2006}
T.~Lee, P.~F. Fischer, {Eliminating parasitic currents in the lattice Boltzmann
  equation method for nonideal gases}, Physical Review E 74~(4) (2006) 046709.
\newblock \href {http://dx.doi.org/10.1103/PhysRevE.74.046709}
  {\path{doi:10.1103/PhysRevE.74.046709}}.

\bibitem{Guo2011}
Z.~Guo, C.~Zheng, B.~Shi, {Force imbalance in lattice Boltzmann equation for
  two-phase flows}, Physical Review E 83~(3) (2011) 036707.
\newblock \href {http://dx.doi.org/10.1103/PhysRevE.83.036707}
  {\path{doi:10.1103/PhysRevE.83.036707}}.

\bibitem{Lou2015}
Q.~Lou, Z.~Guo, {Interface-capturing lattice Boltzmann equation model for
  two-phase flows}, Physical Review E 91~(1) (2015) 13302.
\newblock \href {http://dx.doi.org/10.1103/PhysRevE.91.013302}
  {\path{doi:10.1103/PhysRevE.91.013302}}.

\bibitem{succi2001lattice}
S.Succi, The Lattice Boltzmann Equation for Fluid Dynamics and Beyond, Oxford
  University Press, New York, 2001.

\bibitem{Frisch1986}
U.~Frisch, B.~Hasslacher, Y.~Pomeau, {Lattice-gas automata for the
  Navier-Stokes equation}, Physical Review Letters 56~(14) (1986) 1505--1508.
\newblock \href {http://dx.doi.org/10.1103/PhysRevLett.56.1505}
  {\path{doi:10.1103/PhysRevLett.56.1505}}.

\bibitem{He1997}
X.~He, L.~S. Luo, {Theory of the lattice Boltzmann method: From the Boltzmann
  equation to the lattice Boltzmann equation}, Physical Review E 55~(6) (1997)
  6811--6820.
\newblock \href {http://dx.doi.org/10.1103/PhysRevE.56.6811}
  {\path{doi:10.1103/PhysRevE.56.6811}}.

\bibitem{He1997a}
X.~He, L.~S. Luo, {A priori derivation of the lattice Boltzmann equation},
  Physical Review E 55~(6) (1997) 6333--6336.
\newblock \href {http://dx.doi.org/10.1103/PhysRevE.55.R6333}
  {\path{doi:10.1103/PhysRevE.55.R6333}}.

\bibitem{Chen1991}
S.~Chen, H.~Chen, D.~Martnez, W.~Matthaeus, {Lattice Boltzmann model for
  simulation of magnetohydrodynamics}, Physical Review Letters 67~(27) (1991)
  3776--3779.
\newblock \href {http://dx.doi.org/10.1103/PhysRevLett.67.3776}
  {\path{doi:10.1103/PhysRevLett.67.3776}}.

\bibitem{Chen1992}
H.~Chen, S.~Chen, W.~H. Matthaeus, {Recovery of the Navier-Stokes equations
  using a lattice-gas Boltzmann method}, Physical Review A 45~(8) (1992)
  5339--5342.
\newblock \href {http://dx.doi.org/10.1103/PhysRevA.45.R5339}
  {\path{doi:10.1103/PhysRevA.45.R5339}}.

\bibitem{Qian1992}
Y.~H. Qian, D.~D'Humi{\`{e}}res, P.~Lallemand, {Lattice BGK models for
  Navier-Stokes equation}, Europhysics Letters 17~(6) (1992) 479--484.
\newblock \href {http://dx.doi.org/10.1209/0295-5075/17/6/001}
  {\path{doi:10.1209/0295-5075/17/6/001}}.

\bibitem{Karlin1999}
I.~V. Karlin, A.~Ferrante, H.~C. {\"{O}}ttinger, {Perfect entropy functions of
  the Lattice Boltzmann method}, Europhysics Letters 47~(2) (1999) 182--188.
\newblock \href {http://dx.doi.org/10.1209/epl/i1999-00370-1}
  {\path{doi:10.1209/epl/i1999-00370-1}}.

\bibitem{Ginzburg2008}
I.~Ginzburg, F.~Verhaeghe, D.~D'Humi{\`{e}}res, {Two-relaxation-time lattice
  Boltzmann scheme: about parametrization, velocity, pressure and mixed
  boundary conditions}, Communications in Computational Physics 3~(2) (2008)
  427--478.

\bibitem{Ginzburg2008a}
I.~Ginzburg, F.~Verhaeghe, D.~D'Humi{\`{e}}res, {Study of simple hydrodynamic
  solutions with the two-relaxation-times lattice Boltzmann scheme},
  Communications in Computational Physics 3~(3) (2008) 519--581.

\bibitem{Lallemand2000}
P.~Lallemand, L.~S. Luo, {Theory of the lattice Boltzmann method: Dispersion,
  dissipation, isotropy, Galilean invariance, and stability}, Physical Review E
  61~(6) (2000) 6546--6562.
\newblock \href {http://dx.doi.org/10.1103/PhysRevE.61.6546}
  {\path{doi:10.1103/PhysRevE.61.6546}}.

\bibitem{Lallemand2003}
P.~Lallemand, L.~S. Luo, {Theory of the lattice Boltzmann method: Acoustic and
  thermal properties in two and three dimensions}, Physical Review E 68~(3)
  (2003) 036706.
\newblock \href {http://dx.doi.org/10.1103/PhysRevE.68.036706}
  {\path{doi:10.1103/PhysRevE.68.036706}}.

\bibitem{Luo2011}
L.~S. Luo, W.~Liao, X.~Chen, Y.~Peng, W.~Zhang, {Numerics of the lattice
  Boltzmann method: Effects of collision models on the lattice Boltzmann
  simulations}, Physical Review E 83~(5) (2011) 056710.
\newblock \href {http://dx.doi.org/10.1103/PhysRevE.83.056710}
  {\path{doi:10.1103/PhysRevE.83.056710}}.

\bibitem{McCracken2005}
M.~E. McCracken, J.~Abraham, {Multiple-relaxation-time lattice-Boltzmann model
  for multiphase flow}, Physical Review E 71~(3) (2005) 036701.
\newblock \href {http://dx.doi.org/10.1103/PhysRevE.71.036701}
  {\path{doi:10.1103/PhysRevE.71.036701}}.

\bibitem{Koelman1991}
J.~M. V.~A. Koelman, {A simple lattice Boltzmann scheme for Navier-Stokes fluid
  flow}, Europhysics Letters 15~(6) (1991) 603--607.
\newblock \href {http://dx.doi.org/10.1209/0295-5075/15/6/007}
  {\path{doi:10.1209/0295-5075/15/6/007}}.

\bibitem{Guo2002}
Z.~Guo, C.~Zheng, B.~Shi, {Discrete lattice effects on the forcing term in the
  lattice Boltzmann method}, Physical Review E 65~(4) (2002) 046308.
\newblock \href {http://dx.doi.org/10.1103/PhysRevE.65.046308}
  {\path{doi:10.1103/PhysRevE.65.046308}}.

\bibitem{Kupershtokh2004a}
A.~L. Kupershtokh, {New method of incorporating a body force term into the
  lattice Boltzmann equation}, in: Proceeding of the 5th International EHD
  Workshop, University of Poitiers, Poitiers, France, 2004, pp. 241--246.

\bibitem{Kupershtokh2004b}
A.~L. Kupershtokh, {Incorporating a body force term into the lattice Boltzmann
  equation}, Vestnik NGU (Quarterly Journal of Novosibirsk State Univ.),
  Series: Math., Mech. and Informatics 4~(2) (2004) 75--96.

\bibitem{Kupershtokh2009}
A.~L. Kupershtokh, D.~A. Medvedev, D.~I. Karpov, {On equations of state in a
  lattice Boltzmann method}, Computers and Mathematics with Applications 58~(5)
  (2009) 965--974.
\newblock \href {http://dx.doi.org/10.1016/j.camwa.2009.02.024}
  {\path{doi:10.1016/j.camwa.2009.02.024}}.

\bibitem{Kupershtokh2010}
A.~L. Kupershtokh, {Criterion of numerical instability of liquid state in LBE
  simulations}, Computers and Mathematics with Applications 59~(7) (2010)
  2236--2245.
\newblock \href {http://dx.doi.org/10.1016/j.camwa.2009.08.058}
  {\path{doi:10.1016/j.camwa.2009.08.058}}.

\bibitem{Ginzbourg1994}
{I. Ginzbourg}, {P. M. Adler}, {Boundary flow condition analysis for the
  three-dimensional lattice Boltzmann model}, J. Phys. II France 4~(2) (1994)
  191--214.
\newblock \href {http://dx.doi.org/10.1051/jp2:1994123}
  {\path{doi:10.1051/jp2:1994123}}.

\bibitem{Jamet2002}
D.~Jamet, D.~Torres, J.~U. Brackbill, {On the theory and computation of surface
  tension: The elimination of parasitic currents through energy conservation in
  the second-gradient method}, Journal of Computational Physics 182~(1) (2002)
  262--276.
\newblock \href {http://dx.doi.org/10.1006/jcph.2002.7165}
  {\path{doi:10.1006/jcph.2002.7165}}.

\bibitem{rowlinson2013molecular}
J.~S. Rowlinson, B.~Widom, Molecular Theory of Capillarity, Clarendon Press,
  Oxford, 1982.

\bibitem{Swift1995}
M.~R. Swift, W.~R. Osborn, J.~M. Yeomans, {Lattice Boltzmann simulation of
  nonideal fluids}, Physical Review Letters 75~(5) (1995) 830--833.
\newblock \href {http://dx.doi.org/10.1103/PhysRevLett.75.830}
  {\path{doi:10.1103/PhysRevLett.75.830}}.

\bibitem{Wen2015}
B.~Wen, Z.~Qin, C.~Zhang, H.~Fang, {Thermodynamic-consistent lattice Boltzmann
  model for nonideal fluids}, Europhysics Letters 112~(4) (2015) 44002.
\newblock \href {http://dx.doi.org/10.1209/0295-5075/112/44002}
  {\path{doi:10.1209/0295-5075/112/44002}}.

\bibitem{Zheng2006}
H.~W. Zheng, C.~Shu, Y.~T. Chew, {A lattice Boltzmann model for multiphase
  flows with large density ratio}, Journal of Computational Physics 218~(1)
  (2006) 353--371.
\newblock \href {http://dx.doi.org/10.1016/j.jcp.2006.02.015}
  {\path{doi:10.1016/j.jcp.2006.02.015}}.

\bibitem{Wen2017}
B.~Wen, X.~Zhou, B.~He, C.~Zhang, H.~Fang, {Chemical-potential-based lattice
  Boltzmann method for nonideal fluids}, Physical Review E 95~(6) (2017)
  063305.
\newblock \href {http://dx.doi.org/10.1103/PhysRevE.95.063305}
  {\path{doi:10.1103/PhysRevE.95.063305}}.

\bibitem{Wen2020}
B.~Wen, L.~Zhao, W.~Qiu, Y.~Ye, X.~Shan, {Chemical-potential multiphase lattice
  Boltzmann method with superlarge density ratios}, Physical Review E 102~(1)
  (2020) 013303.
\newblock \href {http://dx.doi.org/10.1103/PhysRevE.102.013303}
  {\path{doi:10.1103/PhysRevE.102.013303}}.

\bibitem{Lele1992}
S.~K. Lele, {Compact finite difference schemes with spectral-like resolution},
  Journal of Computational Physics 103~(1) (1992) 16--42.
\newblock \href {http://dx.doi.org/10.1016/0021-9991(92)90324-R}
  {\path{doi:10.1016/0021-9991(92)90324-R}}.

\bibitem{atkinson2008introduction}
K.~E. Atkinson, An Introduction to Numerical Analysis, John Wiley \& Sons, New
  York, 2008.

\bibitem{Lee2005}
T.~Lee, C.~L. Lin, {A stable discretization of the lattice Boltzmann equation
  for simulation of incompressible two-phase flows at high density ratio},
  Journal of Computational Physics 206~(1) (2005) 16--47.
\newblock \href {http://dx.doi.org/10.1016/j.jcp.2004.12.001}
  {\path{doi:10.1016/j.jcp.2004.12.001}}.

\bibitem{Kumar2004}
A.~Kumar, {Isotropic finite-differences}, Journal of Computational Physics
  201~(1) (2004) 109--118.
\newblock \href {http://dx.doi.org/10.1016/j.jcp.2004.05.005}
  {\path{doi:10.1016/j.jcp.2004.05.005}}.

\bibitem{Huang2011}
H.~Huang, M.~Krafczyk, X.~Lu, {Forcing term in single-phase and Shan-Chen-type
  multiphase lattice Boltzmann models}, Physical Review E 84~(4) (2011) 046710.
\newblock \href {http://dx.doi.org/10.1103/PhysRevE.84.046710}
  {\path{doi:10.1103/PhysRevE.84.046710}}.

\bibitem{vichnevetsky1982fourier}
R.~Vichnevetsky, J.~B. Bowles, Fourier analysis of numerical approximations of
  hyperbolic equations, SIAM, Philadelphia, 1982.

\bibitem{Li2013}
Q.~Li, K.~H. Luo, X.~J. Li, {Lattice Boltzmann modeling of multiphase flows at
  large density ratio with an improved pseudopotential model}, Physical Review
  E 87~(5) (2013) 053301.
\newblock \href {http://dx.doi.org/10.1103/PhysRevE.87.053301}
  {\path{doi:10.1103/PhysRevE.87.053301}}.

\bibitem{Kruger2017}
T.~Kruger, H.~Kusumaatmaja, A.~Kuzmin, O.~Shardt, S.~Goncalo, E.~M. Viggen, {
  The Lattice Boltzmann Method: Principles and Practice}, {Springer
  International Publishing}, 2017.

\bibitem{SETA2007}
T.~Seta, K.~Okui, {Effects of truncation error of derivative approximation for
  two-phase lattice Boltzmann method}, Journal of Fluid Science and Technology
  2~(1) (2007) 139--151.
\newblock \href {http://dx.doi.org/10.1299/jfst.2.139}
  {\path{doi:10.1299/jfst.2.139}}.

\bibitem{Qin2018}
Z.~Qin, W.~Zhao, Y.~Chen, C.~Zhang, B.~Wen, {A pseudopotential multiphase
  lattice Boltzmann model based on high-order difference}, International
  Journal of Heat and Mass Transfer 127 (2018) 234--243.
\newblock \href {http://dx.doi.org/10.1016/j.ijheatmasstransfer.2018.08.002}
  {\path{doi:10.1016/j.ijheatmasstransfer.2018.08.002}}.

\end{thebibliography}




\end{document}